\documentclass[12pt,preprint]{aastex}
\makeatletter

\usepackage{graphics,epsfig,lscape}

\slugcomment{}
\shorttitle{{\sl Chandra} observations of PSR B1929+10. }
\shortauthors{Misanovic, Pavlov,  \& Garmire}

\makeatother

\begin{document}

\title{{\sl Chandra} observations of the pulsar 
B1929+10 and its
 environment}

\author{ Z.\ Misanovic, G.\ G.\ Pavlov, and G.\ P.\ Garmire}

\affil{Dept.\ of Astronomy and Astrophysics,
The Pennsylvania State
University, 525 Davey Lab., University Park,
PA 16802}

\begin{abstract}
We report on  two {\sl Chandra} observations of the 3-Myr-old pulsar B1929+10, 
which reveal a faint compact 
($\sim 9'' \times 
5''$) nebula elongated in the 
direction perpendicular to the pulsar's proper motion, two patchy wings, 
and a possible short ($\sim 3''$) jet emerging from the pulsar. In addition, 
we detect a tail extending up to at least $4'$ in the direction opposite to 
the pulsar's proper motion,
aligned with the $\sim15'$-long tail  detected in  {\sl ROSAT} and 
{\sl XMM-Newton} observations. The overall  morphology of the  nebula  
suggests that the shocked pulsar wind is confined by the ram pressure 
due to the pulsar's supersonic speed. The shape of the compact nebula 
in the immediate vicinity of the pulsar seems to be consistent with the 
current MHD models. However, since these models  do not account yet for 
the change of the flow velocity at larger distances from the pulsar, 
they are not able to constrain the  extent of the long pulsar tail. 
The luminosity of the whole nebula as seen by {\sl Chandra} is $L_{\rm PWN}
\sim 10^{30}$ ergs s$^{-1}$ in the 0.3--8 keV band, for the distance
of 361 pc.
Using the {\sl Chandra} and {\sl XMM-Newton} data, we found that
the pulsar spectrum is comprised of non-thermal (magnetospheric)
and thermal components.
The non-thermal 
 component can be described by a power-law model 
with photon index
$\Gamma\approx 1.7$ 
and luminosity $L_{\rm PSR}^{\rm nonth}\approx 1.7\times 10^{30}$
ergs s$^{-1}$ in the 0.3--10 keV band. The blackbody fit for
the thermal component, which presumably emerges from hot polar caps,
 gives the temperature $kT\approx 0.3$ keV and projected emitting area $A_{\perp}\sim 3\times
10^3$ m$^2$,
corresponding to the bolometric luminosity
$L_{\rm bol}\sim (1$--$2)\times 10^{30}$ ergs s$^{-1}$.
\end{abstract}


\section{Introduction}
\label{intro}

Pulsar wind nebulae (PWNe), generated 
when the magnetized relativistic pulsar winds  
shock in the ambient medium,
 have been observed 
around 
$\approx 50$
pulsars 
\citep[see][for a recent review]{2008AIPC..983..171K}.
The shocked pulsar winds produce  
synchrotron
 radiation  from radio frequencies 
through
 $\gamma$-rays, 
 revealing 
 properties of the winds 
and the ambient medium.

The PWN morphology depends on the pulsar's velocity,  pressure,
 temperature, and magnetic field of the ambient medium, and the geometry of the wind outflow.
For instance,
the torus-jet PWN structures observed  around young pulsars 
in supernova remnants (SNRs), such as the Crab and Vela 
PWNe \citep[]{2000ApJ...536L..81W,2003ApJ...591.1157P},
are commonly interpreted as anisotropic outflows 
with equatorial and polar components
confined by the pressure of the hot gas
in the host SNR interiors, where the speed of sound exceeds the pulsar's speed
 \citep[]{2003MNRAS.344L..93K,2004A&A...421.1063D}.
When the pulsar is moving with a supersonic
speed with respect to the ambient medium, bow-shock PWNe are formed
\citep{2005A&A...434..189B,2005ApJ...630.1020R},
such as the PWNe around PSR J1747$-$2958 
\citep[the Mouse PWN;][]{2004ApJ...616..383G}
 and PSR B1957--20 \citep{2003Sci...299.1372S}.
 The pulsar's supersonic speed causes the ram pressure
 to exceed  the  ambient pressure, so that  the wind termination 
shock acquires  a bullet-like shape \citep[e.g.][]{2005A&A...434..189B}
 and forms a PWN,   sometimes with a long tail   
  extending behind the pulsar. The pulsar's speed becomes supersonic
as it encounters a relatively cold interstellar medium (ISM), 
either after leaving the 
 SNR in which it was born, or when the host SNR dissolves in the ISM
 background. These last stages of the SNR evolution occur
 after $\sim 10^5$ years.

The majority of middle-aged 
($\sim 10^5$--$10^6$ yr) and old ($\gtrsim 10^6$ yr) 
pulsars are expected to  move
 supersonically through the surrounding ISM, and as long as their  winds are
 sufficiently strong, they are expected to produce observable
bow-shock PWNe. Observations of these objects  provide an  
 opportunity to study  bow-shock morphologies and   probe pulsar winds and
 their surroundings at later stages of pulsar evolution.

Until now, only a handful of PWNe around middle-aged pulsars have been
 detected: e.g.,  PSR\,J0538+2817 \citep{2003ApJ...585L..41R}, 
PSR\,B0355+54 \citep{2007Ap&SS.308..309M},
 and Geminga
\citep{2004MmSAI..75..470C,2006ApJ...643.1146P} in X-rays,  
 PSR\,B0906--49  at radio frequencies 
 \citep{1998ApJ...499L..69G}, and 
PSR\,B1951+32 in X-rays and H$\alpha$
 \citep{2004ApJ...610L..33M}.
 However, one of the first bow-shock PWN candidates  
was detected
 serendipitously around a much older pulsar, PSR B1929+10
(hereafter B1929).
This is one of the nearest pulsars 
\citep[$d=361^{+10}_{-8}$ pc;][]{2004ApJ...604..339C},
 with a period $P=226.5$ ms and period derivative
$\dot{P}=1.16\times 10^{-16}$ s s$^{-1}$. Despite the large
spindown age, $\tau\equiv
P/2\dot{P} = 3.1$ Myr, its spindown power, $\dot{E}\equiv 4\pi^2 I\dot{P}/P^3
=3.9\times 10^{33}$ ergs s$^{-1}$ for the moment of inertia $I=1\times 10^{45}$
g cm$^2$, is still relatively high.
As the pulsar's transverse velocity in the plane of the sky,
$V_{\perp}=177^{+4}_{-5}$\,km\,s$^{-1}$ \citep{2004ApJ...604..339C},
substantially exceeds the typical ISM sound speed, one can expect
that the pulsar wind outflow forms a bow-shock PWN. An indication of such a
PWN was first noticed by \citet{1993Natur.364..127W}, who detected
a very long ($\sim1.6$\,pc) tail-like structure 
behind this pulsar in  {\sl ROSAT} data. 
The tail behind B1929
 had been the longest extended structure associated with a compact 
Galactic object until the recent
discovery of a 
6-pc-long X-ray tail behind the middle-aged pulsar 
PSR\,J1509$-$5850, 
reported by \citet{2006HEAD....9.0757K} 
 and described in detail by 
\citet{2008arXiv0802.2963K}
Similar  
(albeit shorter) tails have been found behind a few other pulsars
\citep{2008AIPC..983..171K}
 suggesting that such extended PWN morphologies might be ubiquitous.

The B1929 pulsar  has been extensively observed in X-rays, 
in an attempt to study the magnetospheric and thermal components of its
emission and compare them with the models for pulsar radiation. 
In particular, the thermal emission from 
polar caps heated by relativistic particles accelerated in the pulsar
magnetosphere is predicted by the current pulsar models
\citep[e.g.,][]{2001ApJ...556..987H,2002ApJ...568..862H}, but
 a limited sample of the 
 observed old pulsars
suggests that the nonthermal magnetospheric emission with a power-law spectrum
dominates at higher X-ray energies, $E\gtrsim 2$ keV,
while the nature of radiation at lower energies has been a matter of debate
\citep{2004ApJ...615..908B,2005ApJ...633..367B,2004ApJ...616..452Z,2006ApJ...636..406K}.
Observations of B1929 with the
{\sl Einstein} \citep{1983IAUS..101..471H}, {\sl ROSAT} \citep{1994ApJ...429..832Y,1997A&A...326..682B}, and {\sl ASCA} \citep{1997ApJ...482L.159W,1998AdSpR..21..213K} observatories
 were not able to constrain the nature of the X-ray emission
 from the pulsar
because its spectrum  could be equally well described  as thermal
or  non-thermal. The same result has been obtained by combining the available {\sl ROSAT} and {\sl ASCA}
 data by \citet{2005A&A...434.1097S}, who found that the spectrum of B1929 
could be fitted  either by a power-law or a  two-temperature blackbody 
 model.

Analyzing the images obtained in the 40 ks {\sl ROSAT} PSPC observation,
\citet{1993Natur.364..127W}  
noticed an elongated structure extending approximately 10$'$ behind
 B1929, almost aligned with the direction of its proper motion. 
\citet{1993Natur.364..127W} have suggested that the  detected tail
is synchrotron radiation from the  wind of 
 B1929 confined by the ram pressure.
The detection of this diffuse emission  has also 
 been reported by \citet{1994ApJ...429..832Y} in the analysis of
the same  {\sl ROSAT} PSPC data.  
\citet{1998AdSpR..21..213K} 
also claim detection of some
 diffuse emission in the vicinity of B1929 in the {\sl ASCA} images,
but the poor angular resolution 
of {\sl ASCA} prevented a detailed analysis and did not allow to separate
the contribution from background sources in the alleged PWN emission.

In a multi-wavelength study of B1929 \citep{2006ApJ...645.1421B},
   the tail-like structure extending up to more than 1\,pc behind the pulsar
 has been observed by
 {\sl XMM-Newton}, confirming the {\sl ROSAT} finding.
The spectral analysis suggests that this   
extended emission is nonthermal, likely produced by synchrotron radiation
of the shocked pulsar wind.  
\citet{2006ApJ...645.1421B} also report on an elongated faint diffuse radio
emission found in the Effelsberg radio continuum survey data, whose brightness
distribution roughly coincides with the X-ray tail.
These authors also conclude that the radiation of the B1929 pulsar is 
predominantly nonthermal (i.e., magnetospheric), with a rather soft 
power-law spectrum (photon index $\Gamma\approx 2.7$).

To understand the nature of the extended emission behind B1929
and prove that it is indeed a tail of a bow-shock PWN,
the nebula in the immediate vicinity of the pulsar should be observed with 
an angular resolution  much better than those of {\sl ROSAT} PSPC
($25''$) or {\sl XMM-Newton} ($15''$ 
half-energy width).
Therefore, we conducted two {\sl Chandra}
 observations of the B1929 pulsar and its surroundings. 
The first results, including the detection of a faint nebula in the 
immediate vicinity of
 B1929, have already been reported by \citet{2006HEAD....9.0758M}. 
In this paper we present a more detailed analysis
of these observations.

\section{Observations and results}
\label{observations-and-results}

We observed the field around B1929 with the Advanced CCD Imaging
 Spectrometer (ACIS) aboard 
{\sl Chandra} (Table~\ref{table-observations}).
Both observations were carried
 out in  very faint mode on the ACIS-S3 chip, with the target imaged
about $8''$ from the optical axis.
The data were analyzed
using the {\sl Chandra} Interactive Analysis of Observations (CIAO) software
 (ver.\,3.3.0.1; CALDB ver.\,3.2.0 for observation 6657, and CALDB ver.\,3.2.2 for observation
 7230).

Table~\ref{table-observations} also includes  three archived {\sl
  XMM-Newton} observations of B1929, which we analyzed in addition to the {\sl
  Chandra} data, using SAS (ver. 7.1.0).

\clearpage
\begin{figure}
\begin{center}
\includegraphics[height=5cm,angle=0]{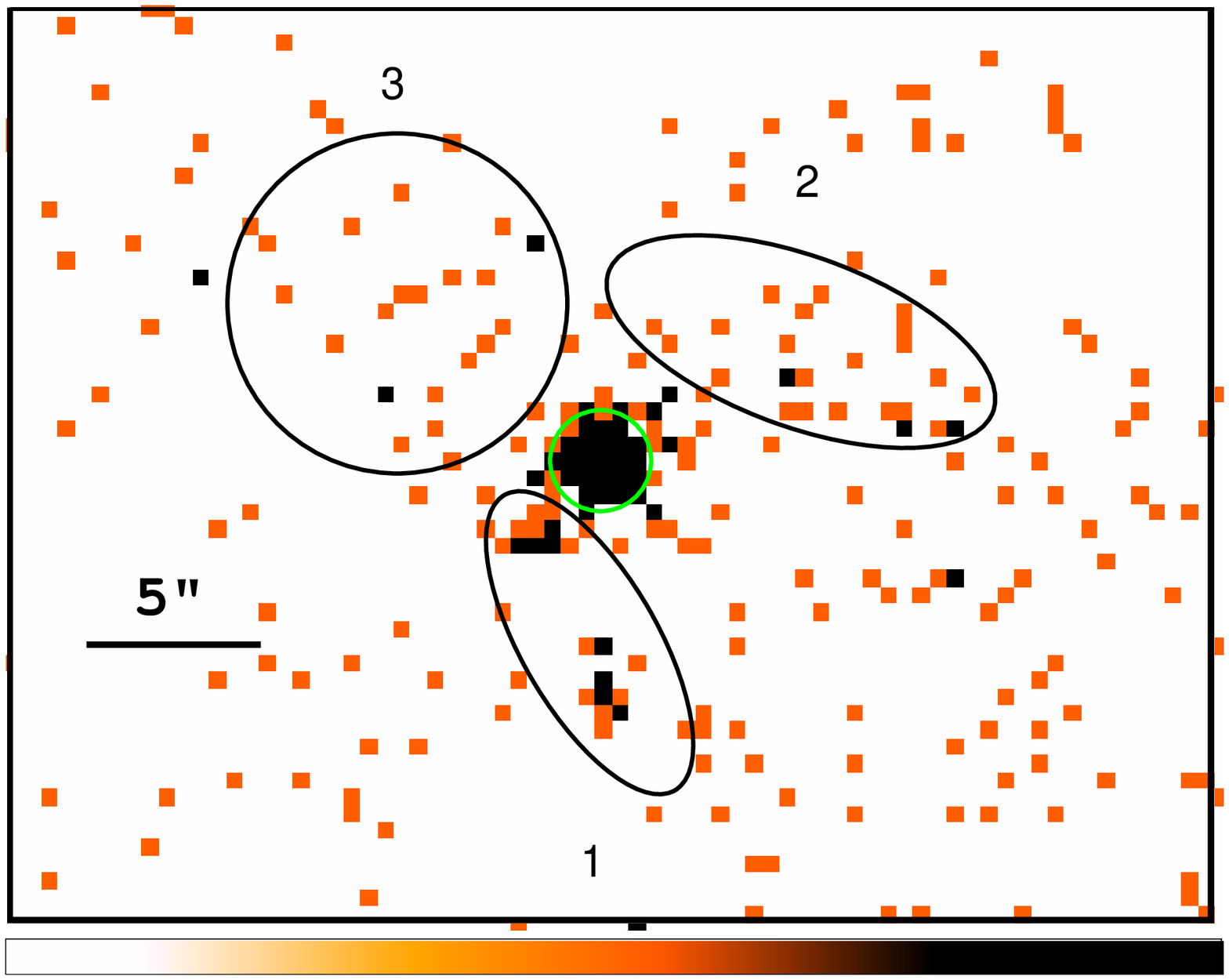}\\
\includegraphics[height=5cm,angle=0]{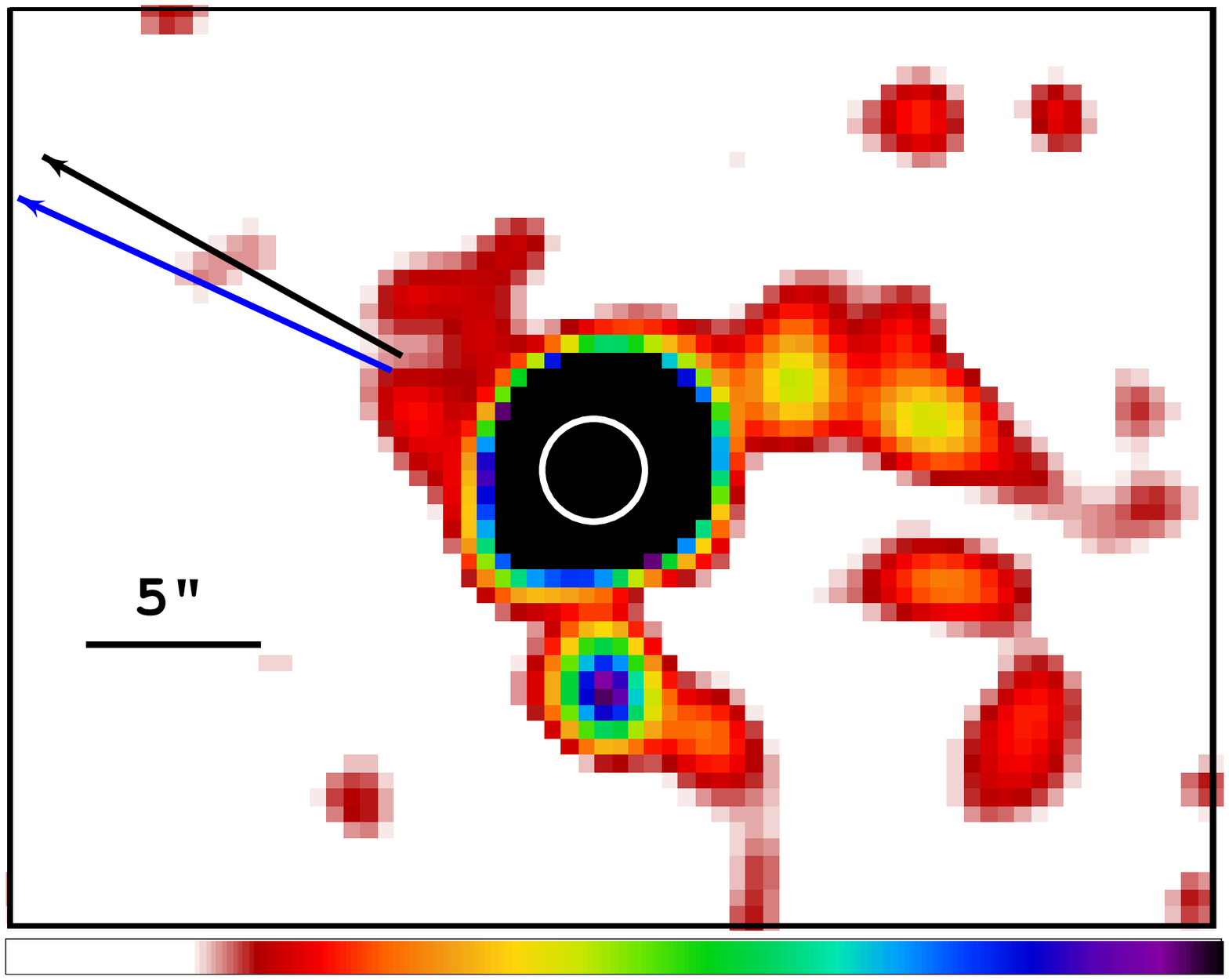}\\
\includegraphics[height=5cm,angle=0]{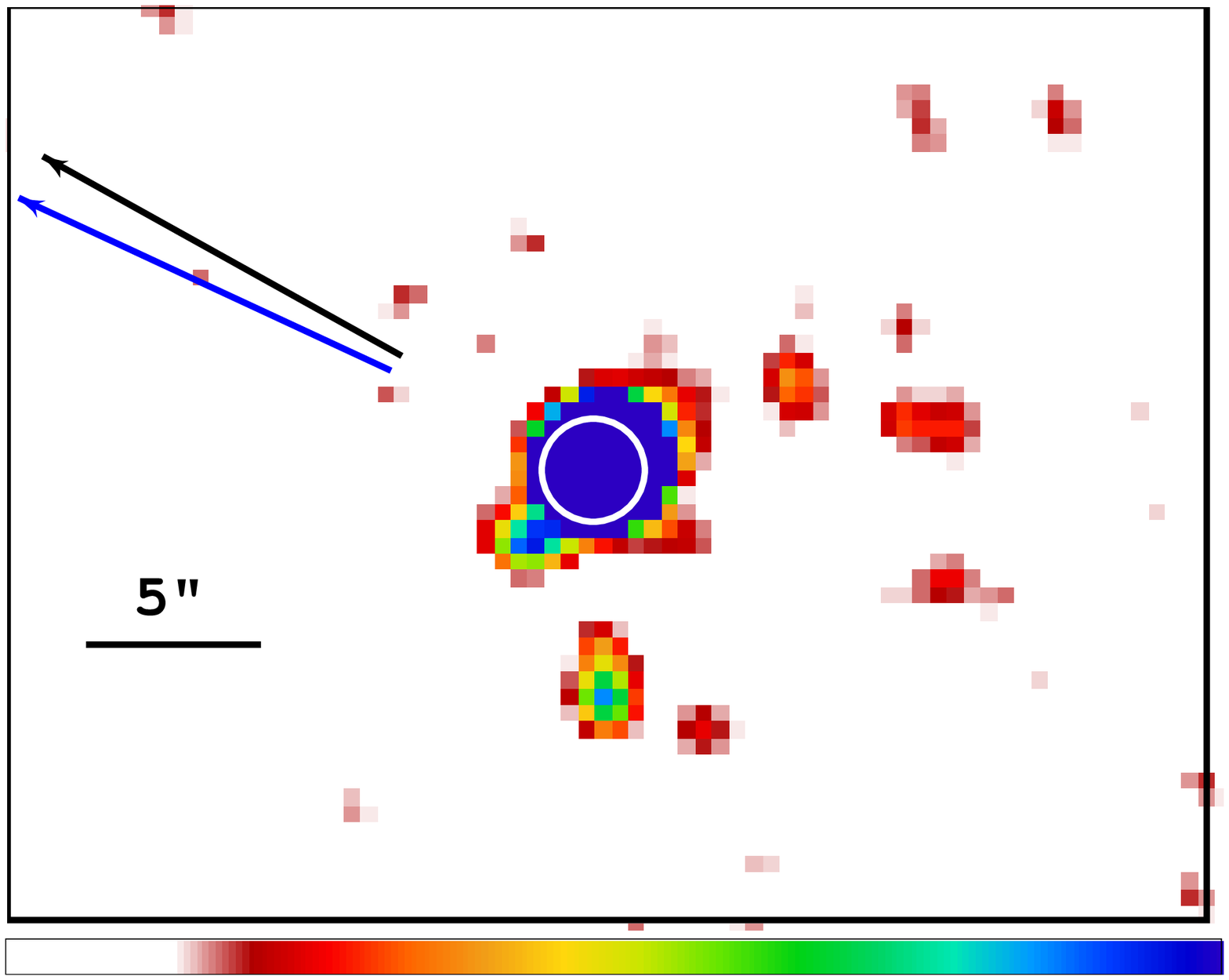}
\end{center}
\caption{ 
Images of the  field around 
B1929 in the 0.3--8 keV band, combining the two ACIS-S3 observations. 
{\em Top}: 
Unsmoothed image (pixel size is $0.492''$).
The circular and  elliptical regions mark the smallest structures detected 
around 
the pulsar at 
$\geq 3 \sigma$ levels
 in at least one of the observations (see Table~\ref{table-regions}).
{\em  Middle}: The same image smoothed using a Gaussian of FWHM $1''$. 
The blue arrow
 shows 
 the direction of the measured pulsar's proper motion
 \citep{2004ApJ...604..339C}, while the black arrow 
indicates the proper motion 
corrected for the Galactic rotation and solar peculiar velocity 
(see text for details).  
{\em Bottom}: 
In this image the pulsar contribution is subtracted (see text for 
details) and smoothing with a Gaussian of FWHM $0.5''$ is applied. 
The brightness scales
in the middle and bottom panels
 are selected to emphasize different components of the extended emission in
 the vicinity of B1929.  
The intensity scale is linear, with a range of 0 to 1.8 
cts/pixel in the top
panel, and 0 to 0.9 
cts/pixel (middle and bottom).
The position of the pulsar  is indicated by
 a 1.5$\arcsec$ circle in all panels.
\label{fig-unsmoothed-images}}
\end{figure}

\clearpage
\begin{figure}
\begin{center}
\includegraphics[height=6.5cm,angle=0]{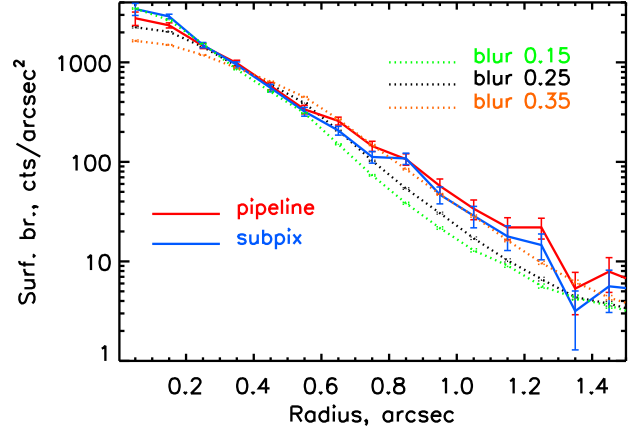}
\caption{ 
  Radial profiles of the combined 
pipeline-processed and subpixelized data (solid lines),
and
MARX-simulated data (dotted lines) for various blurring parameters (see text).
  The counts are measured in annular regions centered at the pulsar's position.
\label{fig-radial-profiles}}
\end{center}
\end{figure} 
\clearpage

\subsection{X-ray imaging}
\label{x-ray-imaging}

We first produced new level-2 event files with the pipeline 
pixel randomization disabled and then applied the
method  by 
\citet{2001ASPC..251..576M} and \citet{2001ApJ...554..496T},
which allows one to reach a subpixel spatial resolution. 
 To minimize the background contribution,
we chose the energy band of 0.3--8 keV.

The only relatively bright source on the S3 chip
is the B1929 pulsar, centered at
R.A. = $19^{\rm h} 32^{\rm m} 13.999^{\rm s}$, 
decl.\ = $10^\circ 59' 32.64''$, and
R.A. = $19^{\rm h} 32^{\rm m} 14.000^{\rm s}$,
decl. = $10^\circ 59' 32.85''$ (J2000)
in the first and second observation, respectively (the centroiding
$1\, \sigma$ uncertainty is
$0.02''$ for each coordinate, 
as determined by the CIAO procedure {\sc celldetect}).
The offsets of these positions
from the radio positions of the pulsar extrapolated to the epochs 
of the two  observations
($0.17''$ and $0.14''$ in R.A., $-0.04''$ and $0.16''$ in decl., 
respectively)
are smaller than the error in the absolute {\sl Chandra} astrometry,
$0.45''$ at the 90\% confidence level for on-axis observations on the S3 
chip\footnote{See \S\,5.4 and
 Fig.\ 5.4 in the Chandra Proposers' Observatory
Guide, ver.\ 10, at http://asc.harvard.edu/proposer/POG.}.

\subsubsection{Combined ACIS image.}

Although both images show
some extended emission in the immediate vicinity of the pulsar,
its surface brightness is very low.
Therefore, to increase the signal-to-noise ratio, 
we have to combine the two ACIS observations.
To align the images, we reprojected the event files to a common position,
corrected the aspect solution of the second observation for the 
pulsar's positional difference of $0.02''$ in R.A. and $0.21''$ in decl., 
and  added the two images.

A nebula surrounding the pulsar 
is clearly seen in the unsmoothed and smoothed images shown in the top and
 middle panels of
 Figure~\ref{fig-unsmoothed-images}. 
To separate this extended
   emission from the pulsar, we simulated the
 ACIS-S3 point source 
observations using the {\sl Chandra} data simulator MARX\footnote{See
http://space.mit.edu/CXC/MARX/}.
We produced simulated images of a point
source at the same position on the detector,
 and with the same X-ray flux and spectral shape
 as those of the observed pulsar,
 running MARX for an effective exposure 100 times longer than the actual
 observations
 to reduce statistical errors.
The simulated images were then  scaled 
to the actual exposure times, combined,
and compared with the data, both pipeline-processed and those
with subpixel resolution.

The width of the simulated point spread function 
(PSF) depends on the value of the MARX parameter
Dither Blur, which is a combination of the aspect reconstruction error, 
ACIS pixelization, and pipeline pixel randomization.
According to the MARX manual\footnote{See also http://www.astro.psu.edu/xray/docs/TARA/ae\_users\_guide/node16.html for a more detailed discussion.},
appropriate values of this parameter are about $0.35''$ if the pixel
randomization is applied and $0.20''$--$0.25''$ if it is switched off, but the
 best values of this parameter may vary from observation to observation.

We found that for the default Dither Blur = $0.35''$,
 the simulated PSF is significantly {\em broader} than
 the core of the observed 
B1929 image, not only for the sharper image with subpixel resolution
 but also for
the pipeline-processed one (see Fig.~\ref{fig-radial-profiles}).
We repeated the simulation for a number of smaller Dither Blur values and found
that  the PSF core of the simulated data with Dither Blur reduced 
to 0\farcs15 matched well to the data with subpixel resolution, 
while a simulation  with Dither Blur = $0.25''$ is close to the 
pipeline-processed
data. 
In both  cases, 
the shape 
 of the simulated  PSF  matched well to the observed one only up to
$\approx 0.6''$ from the center,  
indicating the 
contribution of extended emission at larger distances from the pulsar.
Using the excess of the data counts with respect to the simulation in the
$0.5''$--$1.5''$ annulus, we estimate the signal-to-noise ratio of the PWN
to be $\gtrsim 20$.
For comparison, we also simulated the ACIS-S observation of the central
compact source of the Cas A SNR
(ObsID 6690) 
and found no indication of an extended emission around this point source.

The PWN image in the vicinity of B1929, obtained by
subtracting the simulated data with Dither Blur = $0.15''$ from the 
subpixel resolution image and slight smoothing with a Gaussian of 
$0.5''$ FWHM, 
is shown in the bottom panel of Figure~\ref{fig-unsmoothed-images}.
The image reveals a compact, $\sim 9''\times 5''$, emission 
elongated in the direction perpendicular to the pulsar's proper motion,
two faint wings extending in the direction opposite to the pulsar's motion
and seen up to $\sim 11''$ from the pulsar,
and a hint of a short, $\sim 3''$, linear structure immediately behind the 
pulsar.

\clearpage
\begin{figure*}
\begin{center}
\includegraphics[height=4.7cm,angle=0]{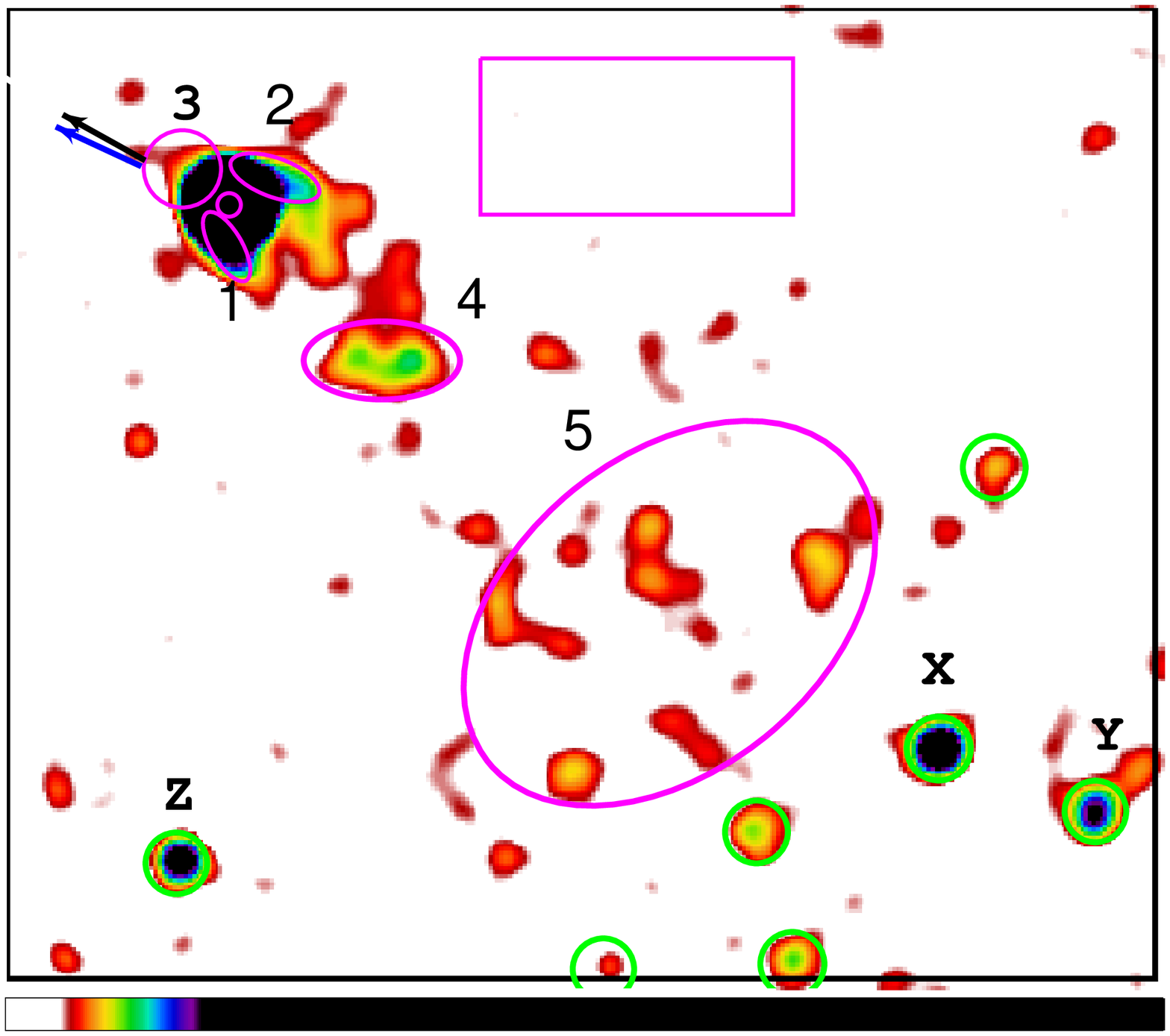}
\includegraphics[height=4.7cm,angle=0]{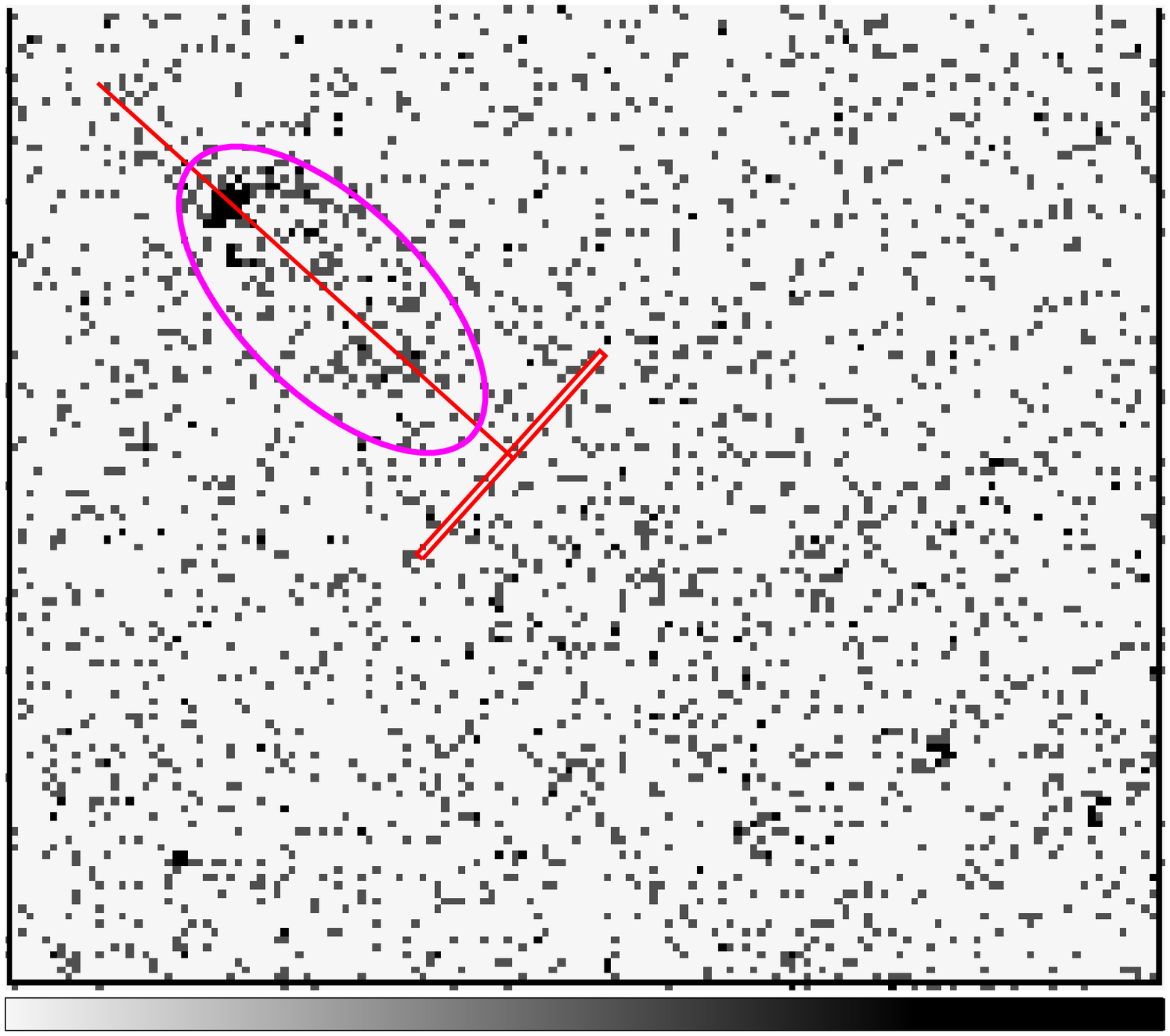}
\includegraphics[height=4.7cm,angle=0]{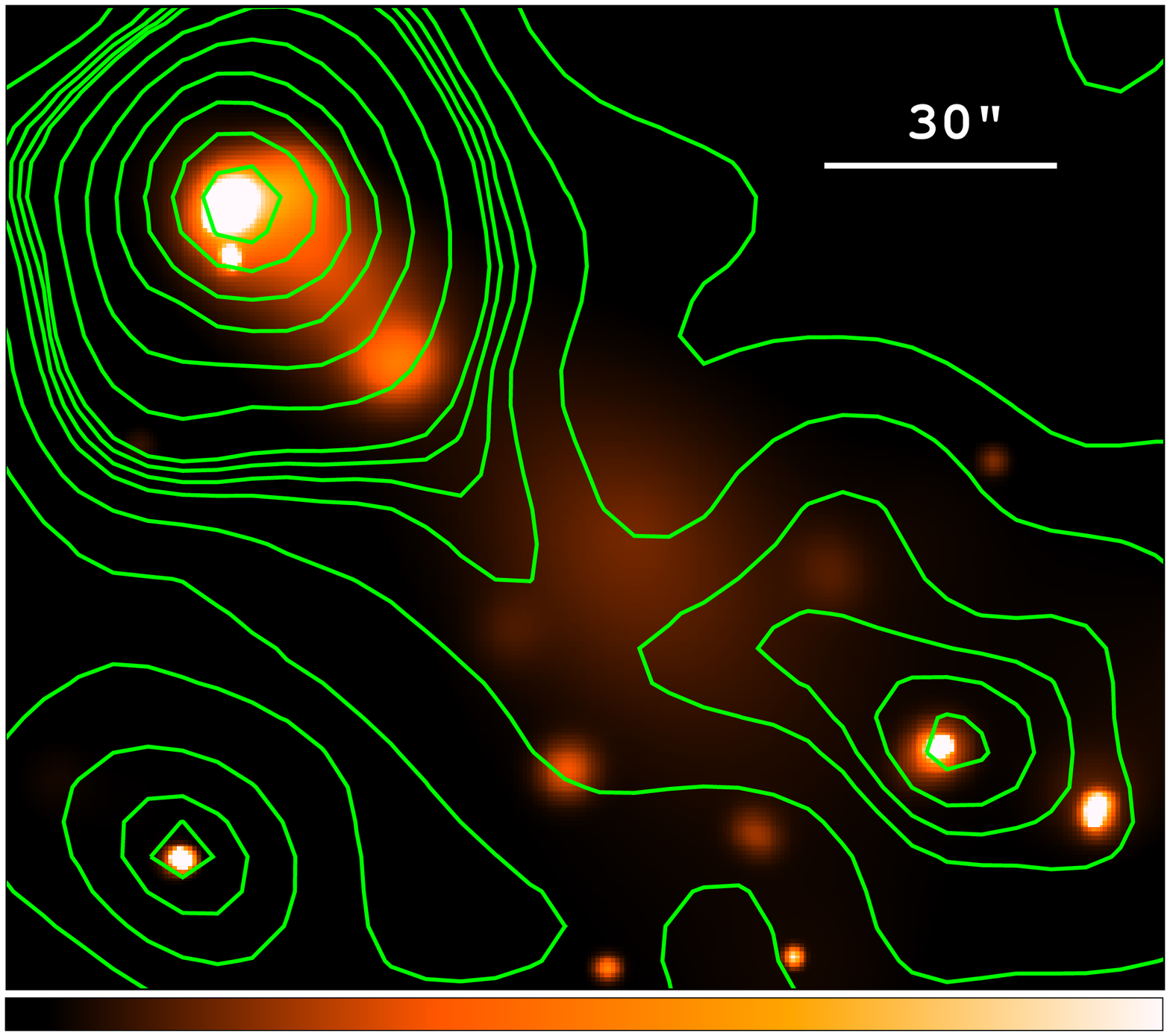}
\end{center}
\caption{
{\em Left:}   
ACIS-S3 image of  
 B1929 and its environment produced by combining observations 6657 and 7230,
in the 0.3--8 keV band. The unbinned  image is smoothed using a Gaussian of FWHM 2\arcsec.
The elliptical regions mark the smallest structures  detected around B1929 
(see Table~\ref{table-regions} and Fig.\ \ref{fig-unsmoothed-images}). 
The pulsar's position is marked by
 a  $1.5''$ circle, while the $40'' \times 20''$ box shows the region 
used for the background evaluation. 
The arrows indicate the direction of the pulsar's proper motion 
(as in Fig.~\ref{fig-unsmoothed-images}).
The green circles enclose the optical/infrared sources in the field of view. The X
marks the X-ray source found with {\sc celldetect} in the combined image,
while Y and Z mark other possible X-ray point sources that might be just below
the decection threshold.
 {\em Middle:} Unsmoothed 
combined image showing the same region. The image is binned by a factor of two. The ellipse
shows the 
 region of enhanced emission, in which the diffuse emission is detected at
 a  $\sim$10$\sigma$ level in the combined image.  The red box and the line
 mark the 
size of the rectangular region and the direction in which the region is moved
 to produce the linear profile shown in Fig.~\ref{fig-profile}. 
 {\em Right:} The same unbinned image smoothed using an adaptive Gaussian kernel with the size self-adjusting
to show the structures with the signal-to-noise ratio in the range 2.2 to 4.
The overlaid contours correspond to the combined {\sl XMM-Newton} observations. The contour levels are the same as in Fig.\ \ref{fig-large}.
 The intensity scale is linear in all  images, with a range of 0 to 0.2
cts/pixel in the
 left and right panels and 0 to 1.8 
cts/pixel (middle). 
\label{fig-combined-conv-smooth-image}}
\end{figure*} 
\clearpage

After confirming that there is a PWN in the immediate vicinity of B1929, 
we proceeded by searching for the nebular emission farther from the pulsar.  
Figure~\ref{fig-combined-conv-smooth-image} shows a $2.5' \times 2.5'$ image
 combining the two {\sl Chandra} observations. The image on the
 left is
 smoothed using a fixed-size Gaussian of 2$\arcsec$ FWHM, while the right panel
 shows a heavily
smoothed image using an adaptive kernel with an adjustable size to show the
 structures with the signal-to-noise ratio in the range 2.2 to 4.
The middle panel shows the combined unsmoothed data binned by a factor of two.
The images
show several regions of extended emission:
 a bow-shock-shaped
 extended emission
 with the Southern Wing (region 1) and Northern Wing (region 2),
 some faint emission in front of the pulsar
 (region 3 or the Front),
faint diffuse emission elongated in the direction
 opposite to the direction of the pulsar's projected velocity, with an enhancement at the
 distance of
 $\approx$\,$30\arcsec$ (region 4 or the Inner Blob).
A more 
extended but fainter region 5 (Outer Blob)
is seen farther behind the pulsar, at a distance of 
 approximately $1.5\arcmin$--$2\arcmin$. 
Regions 1 through 3 are also shown in the top panel of Figure~\ref{fig-unsmoothed-images}, while the numbers of counts in the regions are given in
Table~\ref{table-regions}.

Although these
structures (regions 1 to 5)  
were detected at a level of at least 3$\sigma$ 
 in one or both individual observations, the small number of detected 
counts prevented the detailed spectral analysis.  
Therefore, we selected a larger region
of the extended emission to study the spectral properties of the PWN
   (\S\ref{spectrum-nebula}).
The region is marked by the 
 ellipse ( semi-axes 12$''$ and 25$''$) in the
 middle panel of Figure~\ref{fig-combined-conv-smooth-image}, detected at a level of at least 5$\sigma$ in each
 observation.

The blue arrows in Figures~\ref{fig-unsmoothed-images} 
  and \ref{fig-combined-conv-smooth-image}
 indicate  the direction of the pulsar's proper motion 
 ($\mu _{\alpha}=94.09\pm0.11$\,mas\,yr$^{-1}$, 
$\mu _{\delta}=42.99\pm0.16$\,mas\,yr$^{-1}$)
as measured by  \citet{2004ApJ...604..339C} with respect to extragalactic 
reference radio sources.  
To determine  
the proper motion 
with respect to the local ISM,  
we corrected the
measured proper motion 
 for the 
effects of Galaxy rotation 
and the Sun's peculiar motion with respect to the local standard of rest
 (LSR). 
Following the procedure by \citet{1987AJ.....93..864J}, 
and adopting the value for the Sun's peculiar velocity 
from \citet{1986ApJ...303..724B} and Oort constants from 
\citet{1987gady.book.....B}, 
we obtained the corrected proper motion
$\mu_\alpha=91.6\pm 0.2$ mas yr$^{-1}$,
$\mu_\delta=49.1\pm 0.2$ mas yr$^{-1}$, shown by 
the black arrows in Figures~\ref{fig-unsmoothed-images} and 
\ref{fig-combined-conv-smooth-image}.
The total proper motion, $\mu=104.0\pm 0.3$ mas yr$^{-1}$,
corresponds to the transverse velocity $v_\perp = 178\pm 8$ km s$^{-1}$.
Its direction (position angle $61.7^\circ \pm 0.1^\circ$, 
counted East of North)
 seems to be  
better aligned with the extended tail of diffuse emission,
which further 
supports the PWN interpretation of the detected X-ray emission
 around this pulsar.

Since the observed diffuse emission might be partly due to unresolved 
point X-ray sources in the field of view (e.g., stars or background AGNs), 
we examined
 available
 optical, near-infrared, and radio data of the region around B1929. 
We found several sources (marked with the green circles in Fig.\ \ref{fig-combined-conv-smooth-image}, {\em left}) in the 
USNO-B2 \citep{2003AJ....125..984M} and 2MASS \citep{2003tmc..book.....C}
 catalogs,
 which coincide with the regions
 resembling faint point-like objects in our smoothed images. Although 
  these possible sources are 
very faint in the {\sl Chandra}
 images
(only one 
 of these sources, marked with an X, was detected by the 
CIAO tool {\sc celldetect} above the source threshold of 3, while Y and Z mark
other two possible X-ray point sources that might be just below the detection threshold), 
they might contribute to the observed extended emission.
It is clear, however, that most of the detected tail-like structure 
is indeed diffuse emission associated with the pulsar.

\begin{figure}
\begin{center}
\includegraphics[height=8cm,angle=270]{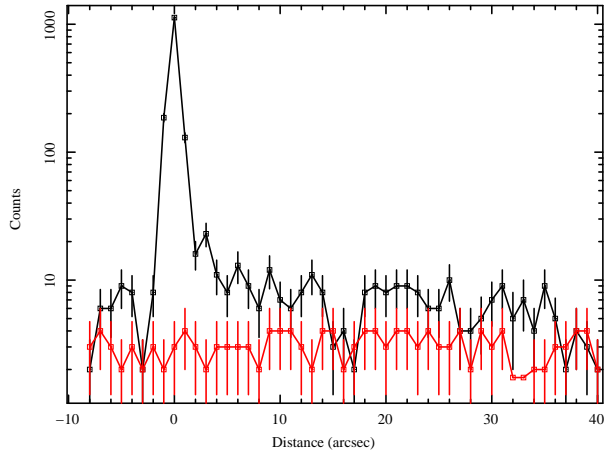}
\end{center}
\caption{X-ray linear brightness profile along the direction of the pulsar's 
tail.
The black data points were obtained
by counting the photons in the $1\arcsec \times 35\arcsec$ 
rectangular regions (see Fig.~\ref{fig-combined-conv-smooth-image}).
The pulsar moves towards the negative values on the X-axis. 
The red  
data points show the X-ray profile of the background emission, 
evaluated in the rectangular regions of the same size on both sides of the
 tail emission.
 \label{fig-profile}}
\end{figure}

Figure~\ref{fig-profile} shows the brightness distribution along the tail, 
which further demonstrates 
the excess emission over the background in front of the pulsar, 
and also in the direction approximately opposite to the projected pulsar's
 velocity. The photons were collected  from  $1\arcsec \times 35\arcsec$
 rectangular regions (see Fig.~\ref{fig-combined-conv-smooth-image}) along the
 tail, and also from the background regions of the same area on both sides of
 the tail in the unsmoothed combined {\sl Chandra} image.
 We show only the brightest
part of the tail, up to $\sim 40''$ from the pulsar, because the extended
 emission farther away is too faint and
cannot be distinguished from the background using such small collecting
 regions, but it is much better seen in the smoothed image.

\clearpage
\clearpage

\begin{figure*}
\begin{center}
\includegraphics[height=5cm,angle=0]{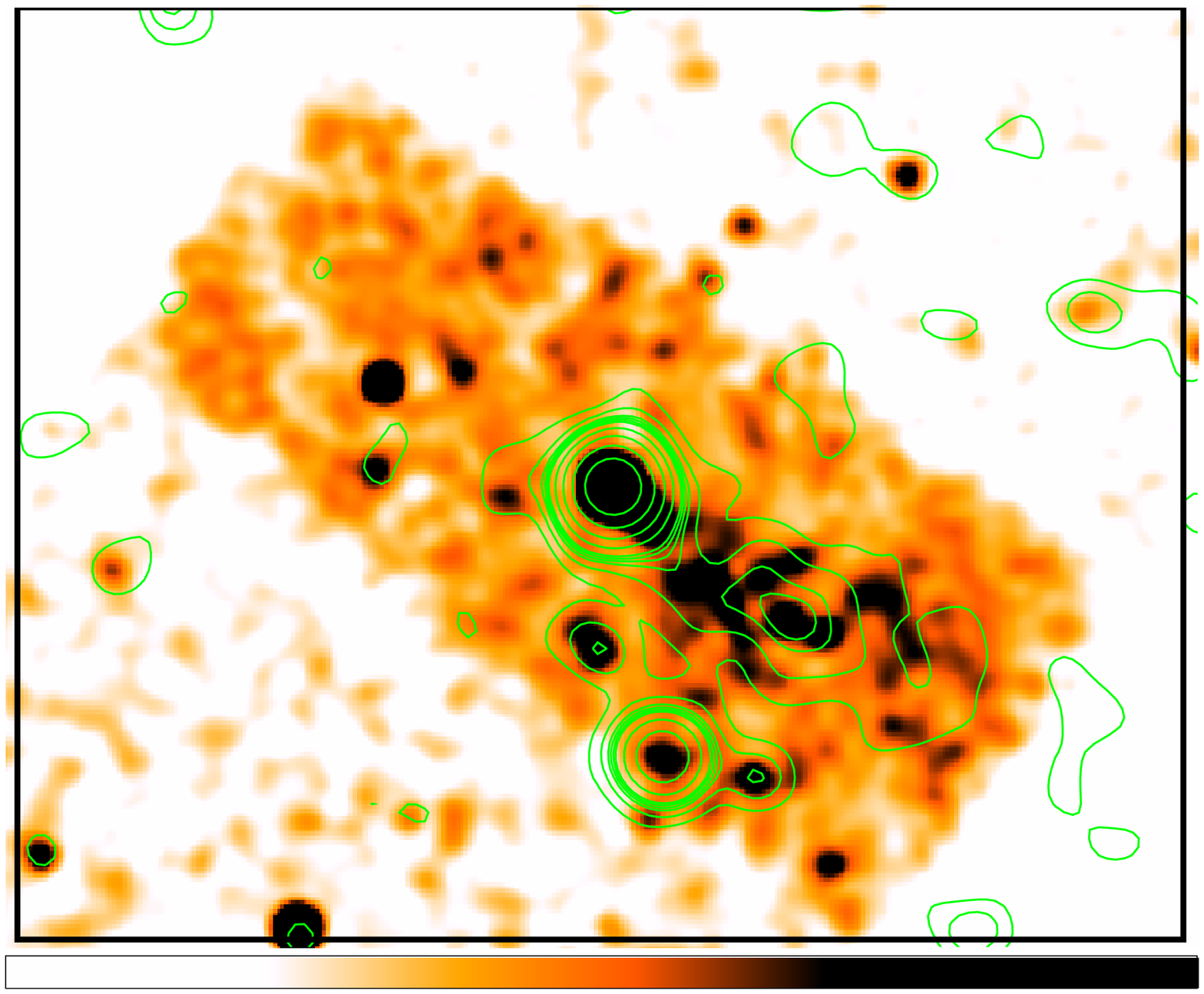}
\includegraphics[height=5cm,angle=0]{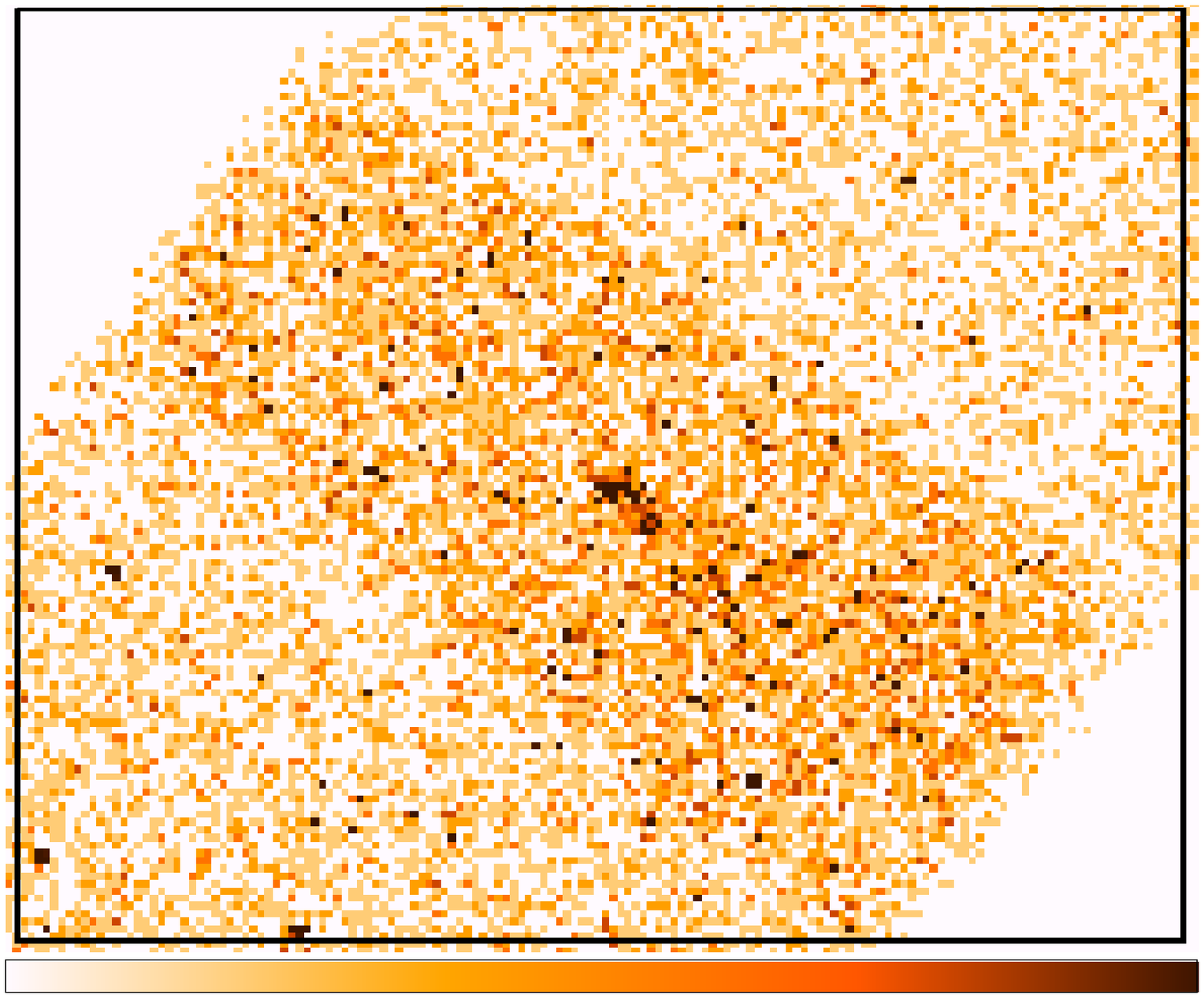}
\end{center}
\caption{
10\arcmin $\times$ 8\arcmin\,  ACIS-S3 field around B1929. 
The image on the left is produced from the combined event file, binned
by a factor of 4 and then smoothed by a Gaussian of FWHM 6$''$.
The contours, at levels 
0.08, 0.1, 0.12, 0.13, 0.14, 0.15, 0.2, 0.3, 0.5, 0.9, 2 and 5 
counts\,pixel$^{-1}$, correspond to the three combined 
{\sl XMM-Newton} observations (MOS1 and MOS2 data, pixel size 1.1$''$).
   The right panel shows the same image, which is binned
 by a factor of 8 but not smoothed.
To minimize the background, both images  include only
the 0.5--7 keV 
 energy range. 
The intensity scale is linear, with ranges of 0 to 0.8 
counts\,pixel$^{-1}$ (left) and
0 to 6 counts\,pixel$^{-1}$ (right).
 \label{fig-large}}
\end{figure*}

To look for extended emission at even larger distances from the
pulsar and compare it with the $10'-15'$ tail detected by
\citet{2006ApJ...645.1421B}, we
produced heavily binned 
images shown in 
Figure~\ref{fig-large}. 
Although 
the smaller ACIS-S field of view and
 the relatively short {\sl Chandra} exposure
do not allow us to detect the full extent of the faint diffuse 
emission observed by {\sl XMM-Newton}, 
at least some parts of the B1929 long tail 
are 
seen up to $\approx 4'$ from the pulsar in  
the combined ACIS-S3 image.
 The long tail extends  from the 
compact PWN resolved in the {\sl Chandra} images in the same direction 
as the emission detected in {\sl XMM-Newton} data.

\clearpage
\clearpage

\subsubsection{Individual observations and variability}
\label{variability}

\begin{figure}
\begin{center}
\vspace{3mm}
\includegraphics[height=5.2cm,angle=0]{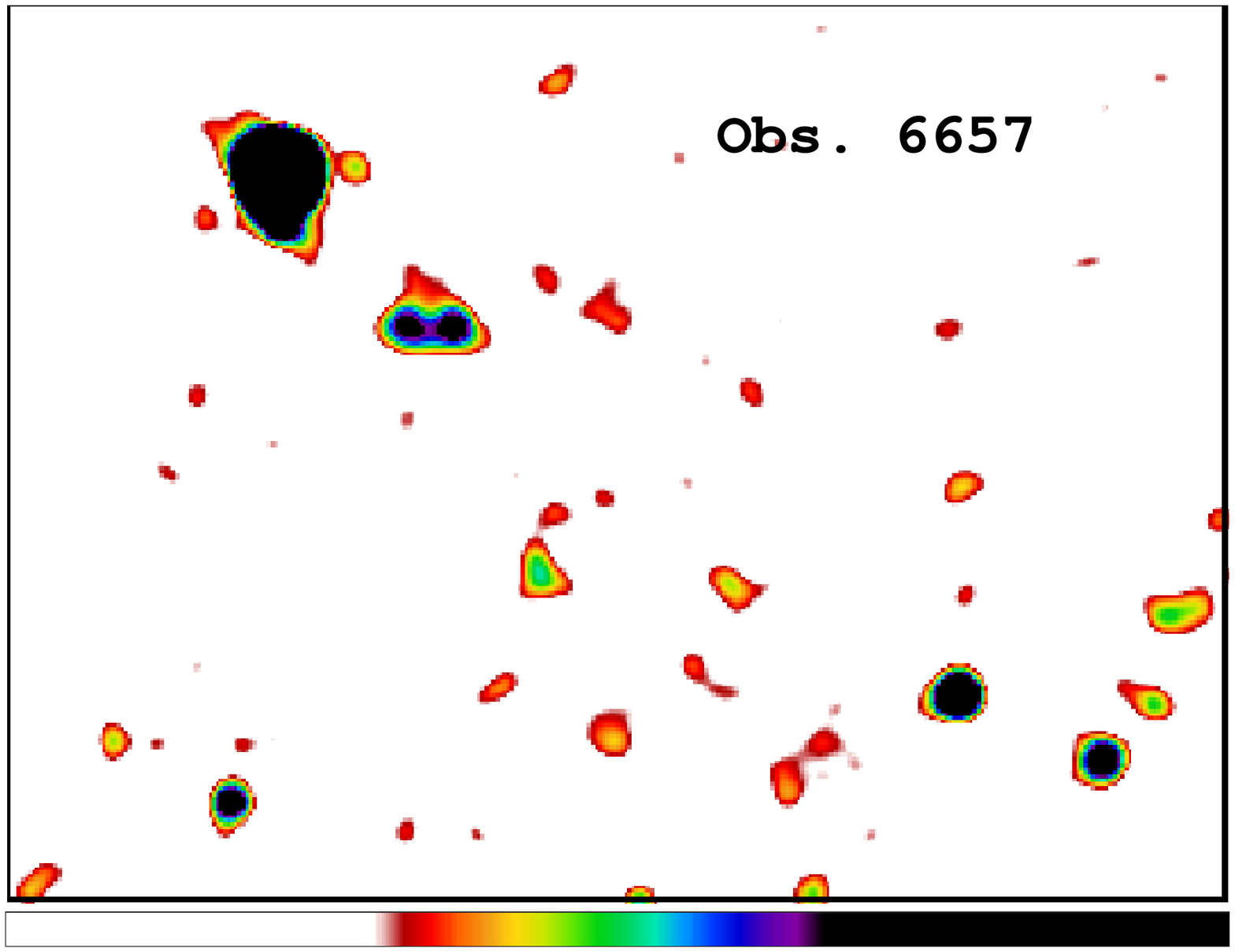}
\includegraphics[height=5.2cm,angle=0]{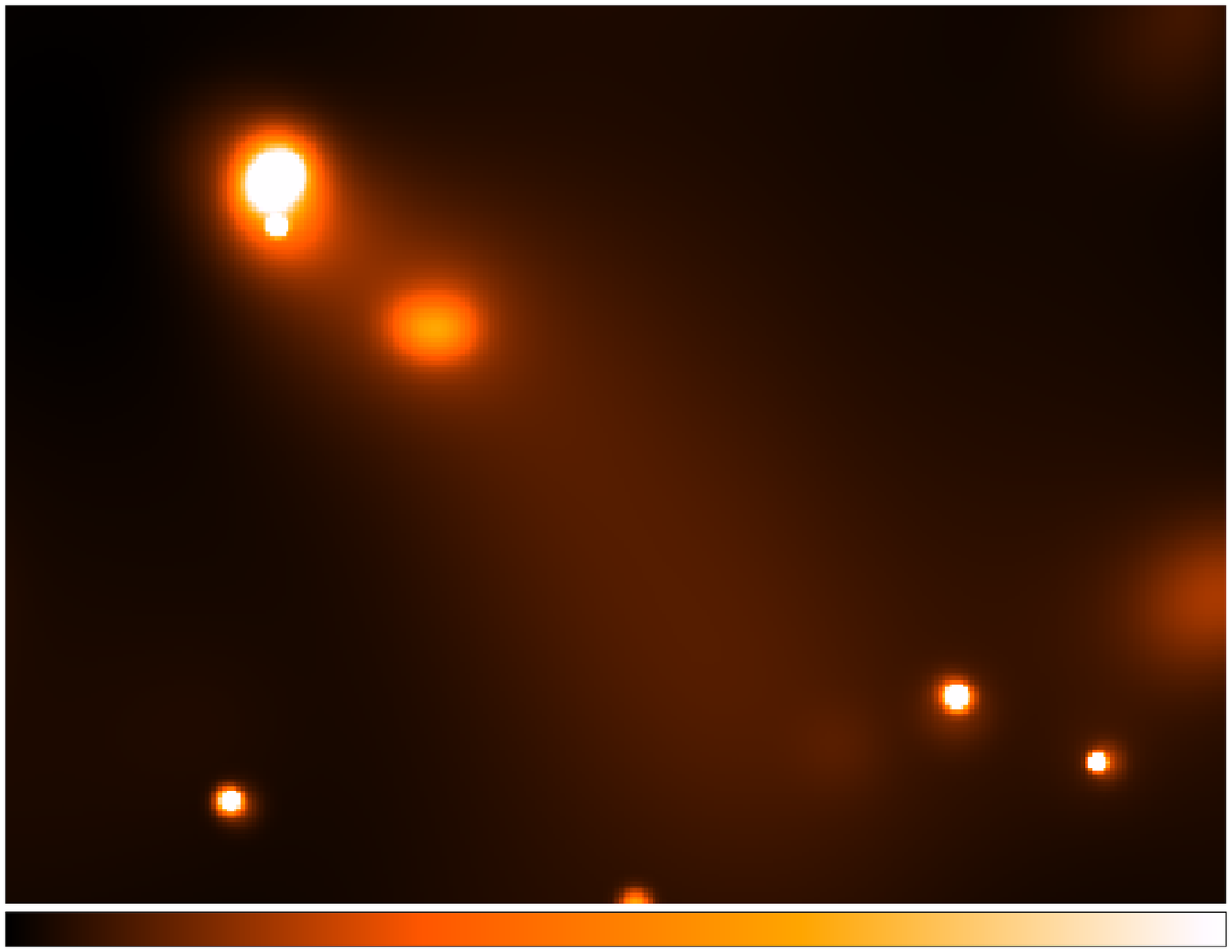}
\includegraphics[height=5.2cm,angle=0]{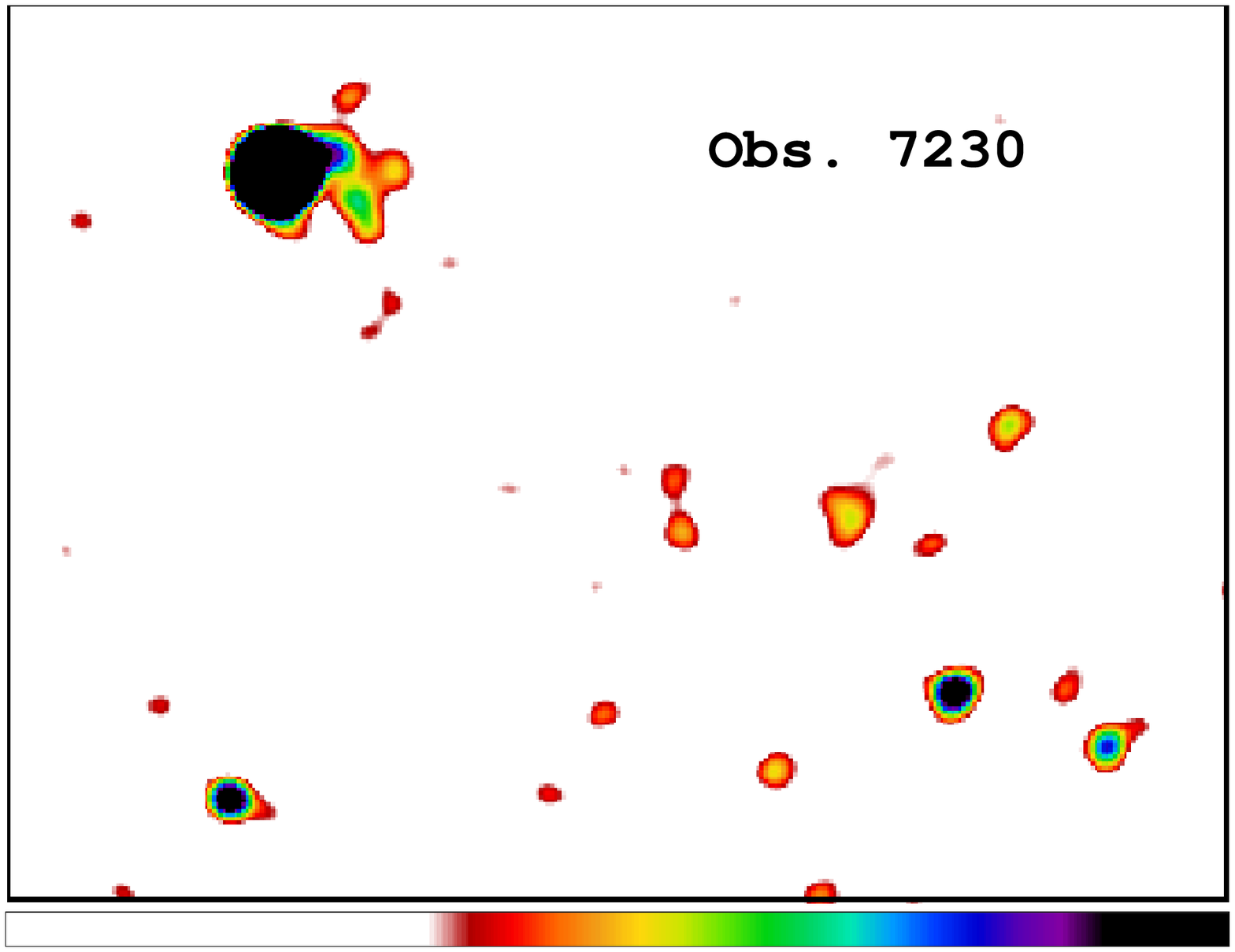}
\includegraphics[height=5.2cm,angle=0]{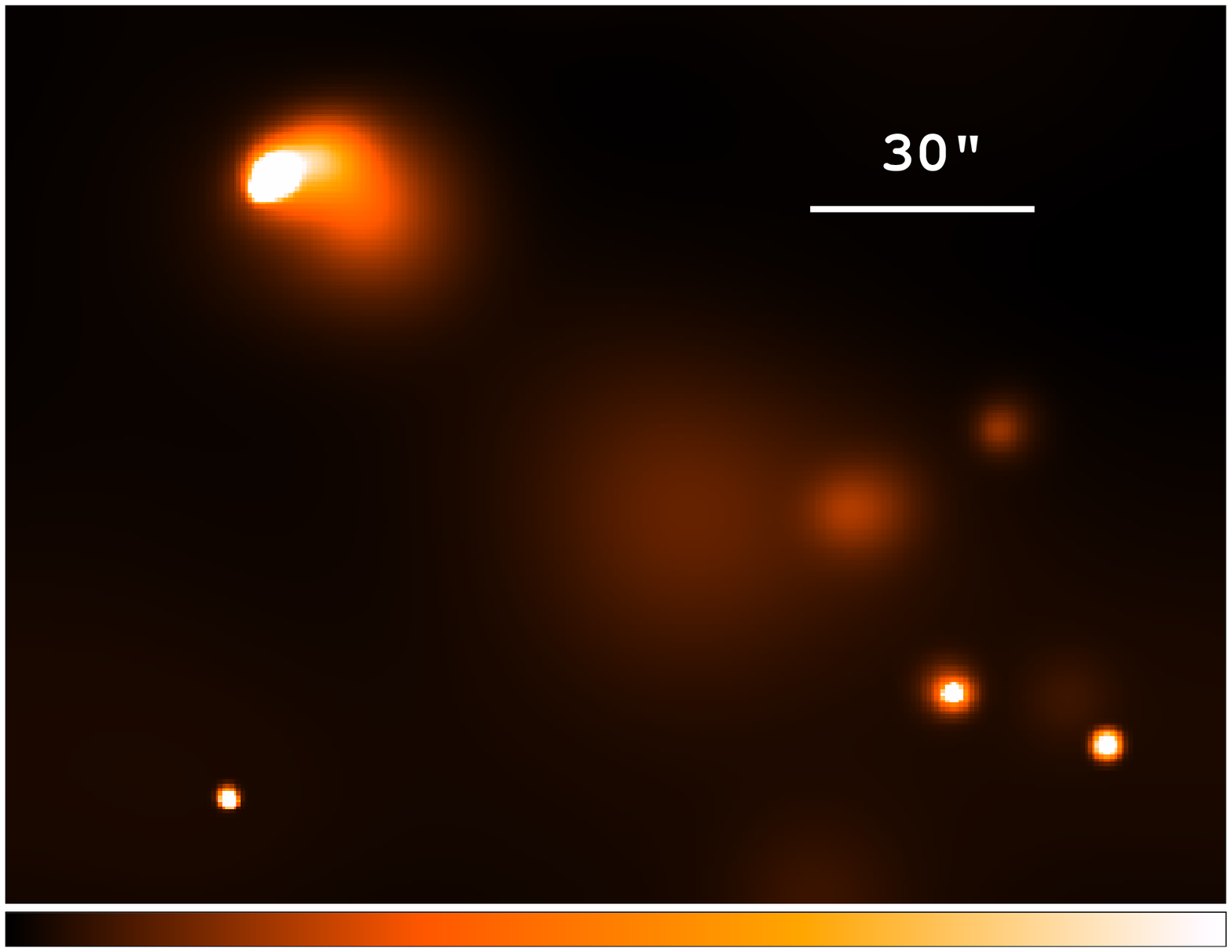}
\end{center}
\caption{ {\em Left}: 
0.3-8\,keV  images of 
the first ({\em top}) and 
second ({\em bottom}) ACIS-S3 observations, 
 smoothed using a Gaussian of FWHM 2\arcsec.
{\em  Right}: The same images adaptively smoothed to show the structures 
with the signal-to-noise ratio in the range 2.2 to 4.
The intensity scale is linear with a range of 0 to 0.2 cts/pixel 
in all panels.
 \label{fig-individual-conv-smooth-image}}
\end{figure}

Since some PWNe show significant temporal variations on time-scales of months 
\citep[e.g., the Crab and Vela PWNe,][]{2002ApJ...577L..49H,2003ApJ...591.1157P}, we have examined the individual ACIS-S observations, which are separated 
by $\simeq$6 months, in search for  variability of the  PWN of B1929.
Figure~\ref{fig-individual-conv-smooth-image} shows the 
images of
 the individual
{\sl Chandra} observations of B1929 and its PWN.
Similarly to the combined data, the images on the left were smoothed 
using a Gaussian of FWHM 2\arcsec, while on the right we show  the same 
images adaptively smoothed to reveal structures with the 
signal-to-noise ratio in the range 2.2 to 4. The bright features seen
in the combined image (Fig.~\ref{fig-combined-conv-smooth-image}) are also
 visible in these individual images, but their apparent brightness seem to 
show a substantial difference between the observations.
For example, the Inner Blob (region 4), clearly detected in the
 observation 6657
(December 2005), is not visible in the 
image of the observation 7230 (May 2006), while the Outer Blob 
(region 5) appears fainter and
slightly smaller in the second observation. 
 In addition,
the Southern Wing and the emission in front of the pulsar (region 3) are
 slightly brighter in the observation 6657, while 
 the Northern Wing is brighter in the second observation.

To quantify this apparent variability,
 we measured the total background-subtracted 
counts detected in each 
of the extended regions 
in the unsmoothed images of both data sets,
 and  calculated the corresponding 
surface brightness 
(listed in Table~\ref{table-regions}). The background
emission was determined using the rectangular region with an area of
 784 arcsec$^2$  shown in
  Figure~\ref{fig-combined-conv-smooth-image} (left), from which we measured 
  $(3.3\pm0.4)\times 10^{-6}$\,cts\,s$^{-1}$\,arcsec$^{-2}$ and $(2.2\pm0.4)\times 10^{-6}$\,cts\,s$^{-1}$\,arcsec$^{-2}$  in the observations 6657 and 
7230, respectively.

As shown in Table~\ref{table-regions}, the
 surface brightness of the five extended regions is found to change 
by a factor of two (e.g. region 1) or  even 
four (region 3). However, due to the large statistical errors,
 the variability 
could only be detected at a level of 
up to $\approx2.5\sigma$ (region 5).
Deeper observations would be needed to further  examine possible variability
 of the nebula associated with this old pulsar.
 We also note that the variation of the pulsar count rates
between the two {\sl Chandra}
observations was found to be within the statistical uncertainty
  (see \S~\ref{spectral-analysis-pulsar}).

\clearpage
\clearpage

\subsection{Connection with the {\sl XMM-Newton} data}
\label{xmm}

To examine the connection between the
{\sl Chandra} and {\sl XMM-Newton} data on the PWN, we produced
 combined
MOS1 and MOS2\footnote{The PN observations were taken in the small-window mode and have high background.}
  broad-band images for each of the three available
 {\sl XMM-Newton}
 observations of the B1929 field
 listed in Table~\ref{table-observations}
 \citep[see][for more detail about the {\sl XMM-Newton} data
and for the combined image of all three observations]{2006ApJ...645.1421B}.

\begin{figure*}
\begin{center}
\includegraphics[height=5cm,angle=0]{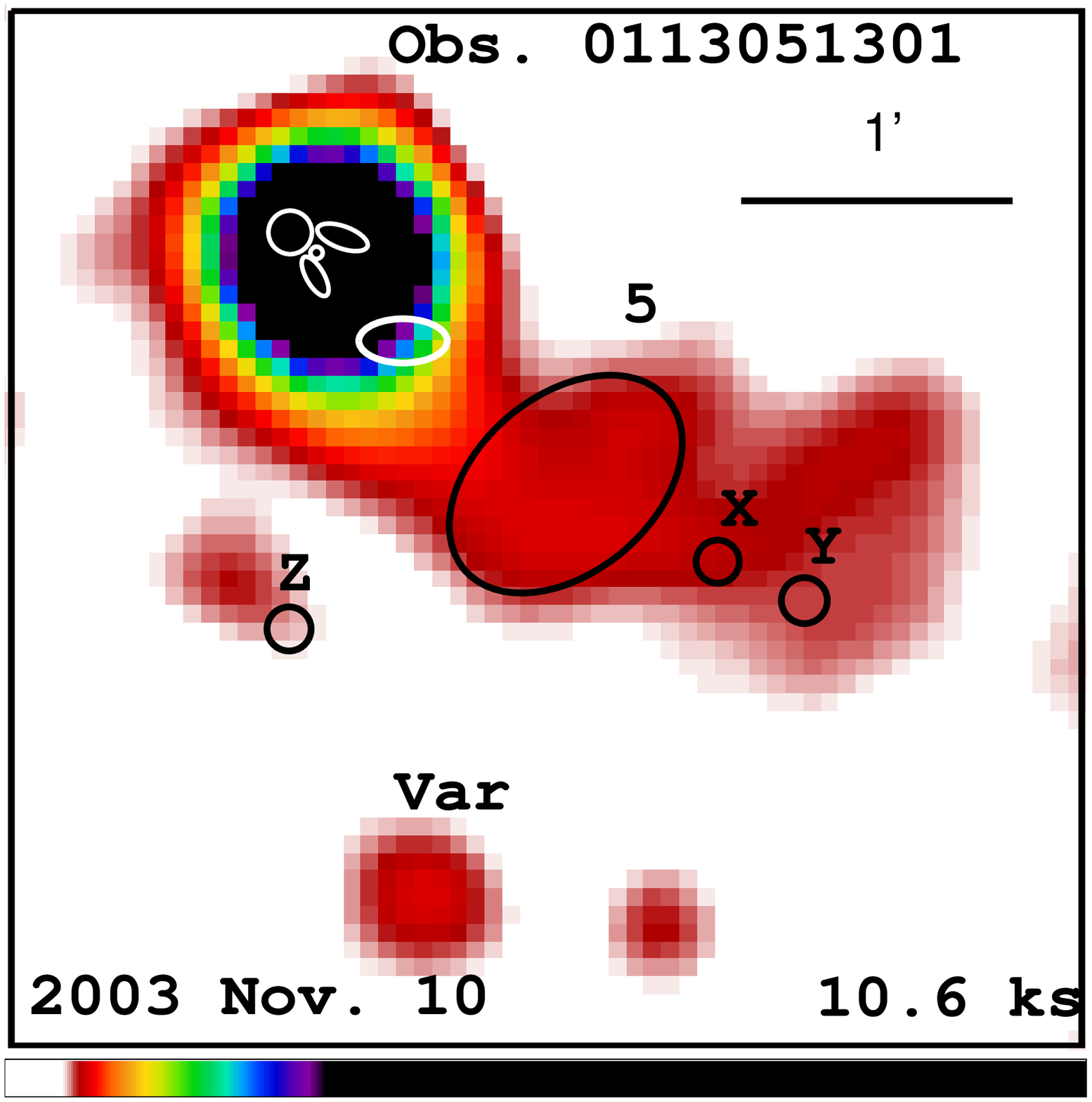}
\includegraphics[height=5cm,angle=0]{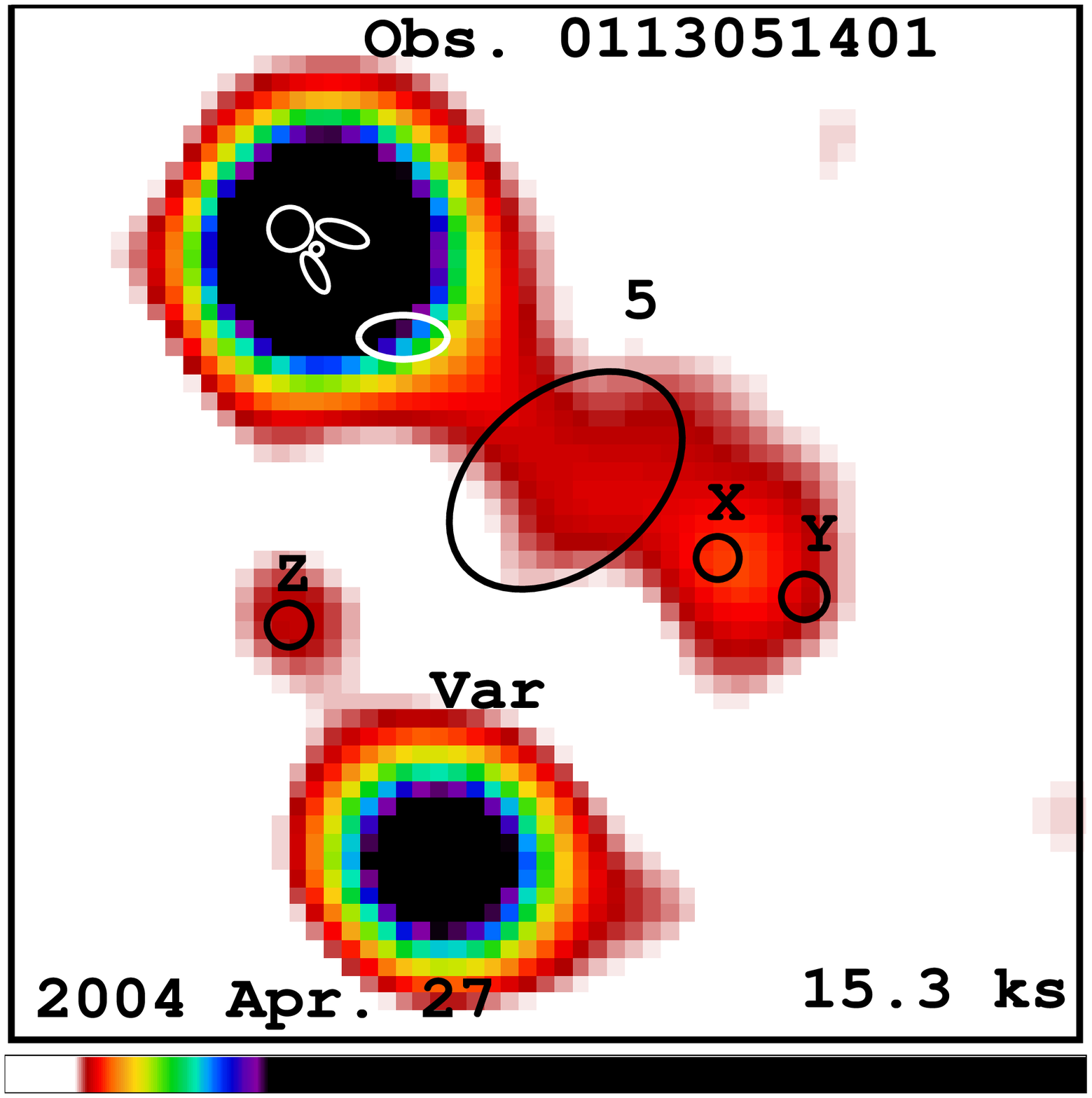}
\includegraphics[height=5cm,angle=0]{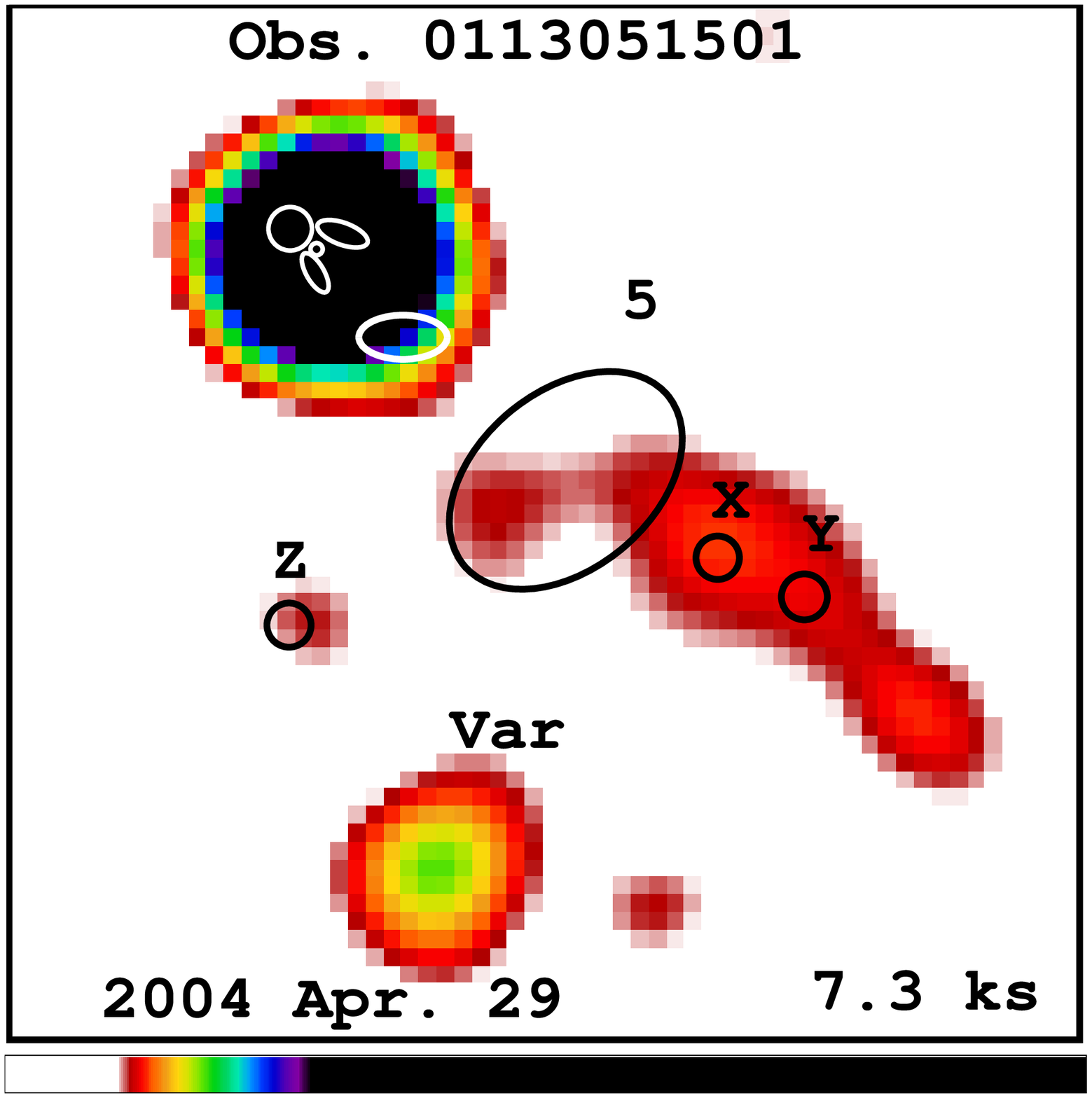}
\end{center}
\caption{Combined MOS1+MOS2 images of B1929 and its surroundings in the three {\sl XMM-Newton} observations.
Broad-band (0.3--10\,keV) images were smoothed using a Gaussian of FWMH $6''$. 
 The five regions of enhanced emission and possible pointlike
 sources X, Y and Z detected in the  ACIS-S observations are
shown (see
left panel of
Fig.\ \ref{fig-combined-conv-smooth-image}
and Table~\ref{table-regions}).
The bright source (Var) $\sim2.5'$
 south from the pulsar
exhibits strong variability, also clearly visible in the unsmoothed MOS
 images.
The intensity scale is linear, with a range of 0 to 6 counts per pixel in all panels.
\label{fig-xmm}}
\end{figure*}

 The smoothed images are shown in Figure~\ref{fig-xmm},
 overlayed with the
structures detected in the
 ACIS-S observations. 
 Since the
 MOS PSF is much broader than that of ACIS,
the compact PWN (regions 1--3 in Figs.~\ref{fig-unsmoothed-images} and \ref{fig-combined-conv-smooth-image}) and the beginning of the
PWN tail (region 4) cannot be resolved from the pulsar.
However, at least in the first two {\sl XMM-Newton} observations
 we clearly see the local brightening in the PWN tail $\approx 2'$
 from the pulsar, approximately coinciding 
with the Outer Blob (region 5) detected
 in the {\sl Chandra} data.
From  the three unsmoothed MOS1 images, we calculated an average background-subtracted count rate of
$1.5 \pm 0.3$\,counts\,ks$^{-1}$ in the Outer Blob.
With account for different ACIS-S and MOS1 responses,
this count rate is consistent
with
that measured in the combined {\sl Chandra} data in the same region
(see Table~\ref{table-regions}).

\clearpage
\clearpage

Since our {\sl Chandra} data  indicate  possible variability of
 the  PWN, we examined the {\sl XMM-Newton} data
for brightness variations in the extended
 emission.
From the number of the background-subtracted counts in the Outer Blob,
we determined the averaged MOS1 count rates of
$1.7 \pm 0.5$, $1.6 \pm 0.5$, and $1.3 \pm 0.5$ counts ks$^{-1}$
in observations 0113051301, 0113051401 and 0113051501, respectively.
Thus, although the Outer Blob changes its surface brightness by a factor of 2 
at a $2.5 \sigma$ level in our
ACIS-S data, the surface brightness
of the same region in the MOS1
data,
 separated  by approximately the same period of $\sim$5 months as the two
 {\sl Chandra} pointings, does not show a significant brightness change.

Figure~\ref{fig-xmm} suggests that
 the  region
southwest of the Outer Blob
might be variable. 
However, in the ACIS image we see no
diffuse emission in this area, 
but there are two faint pointlike sources, X and Y
(see Figs.\ \ref{fig-combined-conv-smooth-image} and \ref{fig-xmm}).
Hence, the apparent variability of this region could be due to the variability 
of these two 
sources, which could not be resolved in the {\sl XMM-Newton} observations.
Thus, although the  {\sl Chandra}  observations
hint that the tail of B1929 might be variable on
  time-scales of several months,    
 longer high-reslolution observations are required
to firmly establish this variability.

Interestingly, the {\sl XMM-Newton} images show
a highly variable
source  (Var) $\sim2.5\arcmin$
south of the pulsar.
It was almost as bright as the pulsar in the 2004 April 27 observation,
 while it is
barely visible in the 2003 November 10 image.
The  source was at the edge of the S3 chip in the first {\sl Chandra}
 observation and barely detectable. However, it completely disappeared in the second observation, although its position was within the field of view.  
As it is apparently a point source, strongly detached from the other
PWN structures, we conclude that it is unrelated to the B1929 PWN.


\clearpage
\begin{figure}
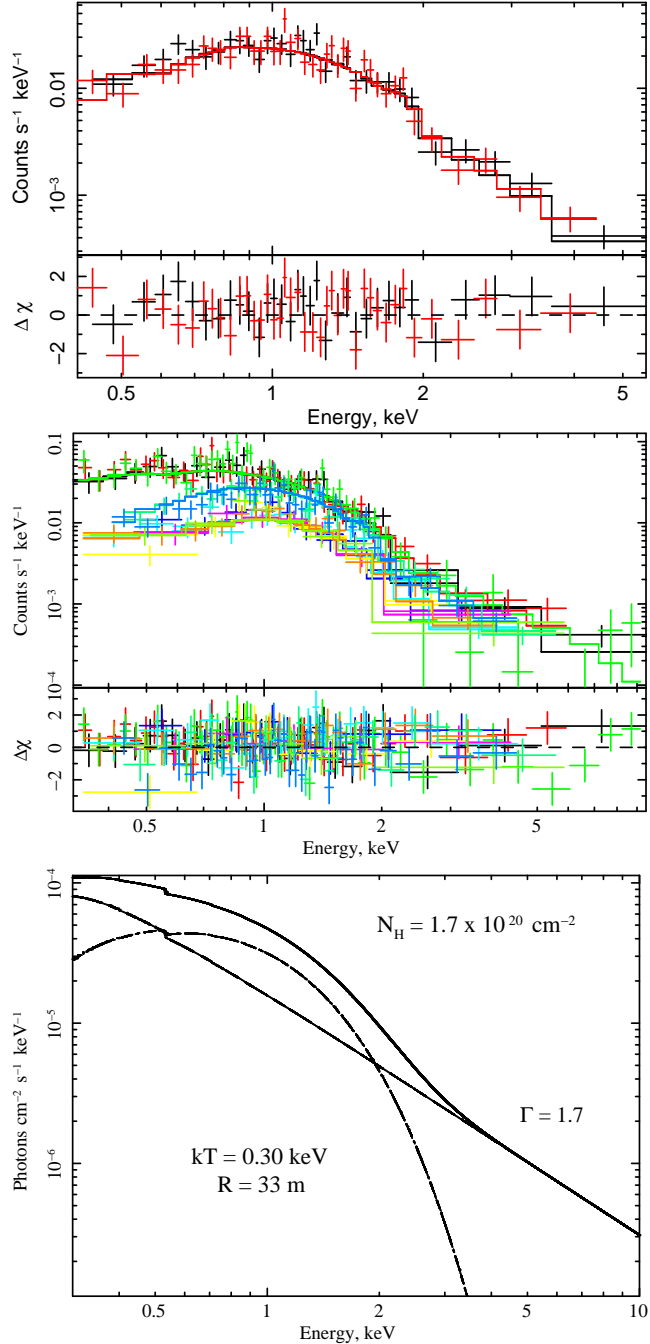

\begin{center}
\includegraphics[height=8.5cm,angle=270]{fig8a.ps}
\includegraphics[height=8.5cm,angle=270]{fig8b.ps}
\includegraphics[height=8.5cm,angle=270]{fig8c.ps}
\end{center}
\caption{{\em Top}: Spectra of B1929 extracted from the 
{\sl Chandra} observations 
6657 (black) and 7230 (red)
and the best-fit absorbed PL model. 
The model parameters are given in Table~\ref{table-spectrum}. 
{\em Middle}: PN, MOS1 and MOS2 spectra of B1929 extracted from the 
three {\sl XMM-Newton} 
observations (PN: black, red, green; MOS1: blue, cyan, magenta; MOS2: yellow, orange, lime-green; in the observations  0113051301, 0113051401 and 0113051501, respectively),
and fitted simultaneously with the {\sl Chandra}
 data (Obs.\ 6657: dark-green; Obs.\ 7230: dark-blue)  using the  
absorbed PL+BB model. The model parameters are given in
Table~\ref{table-spectrum}.
{\em Bottom:} Photon spectrum for the best-fit 
absorbed BB+PL model and its components.
The unabsorbed luminosity of the BB component,
  $\approx$ $1.2 \times 10^{30}$ ergs s$^{-1}$,  is estimated to be
 $\sim$45\% of the total luminosity,
 $\approx$ $2.6  \times 10^{30}$ ergs s$^{-1}$,
 in the 0.3--8\,keV band.
\label{fig-combined-spectra-pl}}
\end{figure}    
\clearpage

\clearpage
\begin{figure}
\begin{center}
\includegraphics[height=8cm,angle=90]{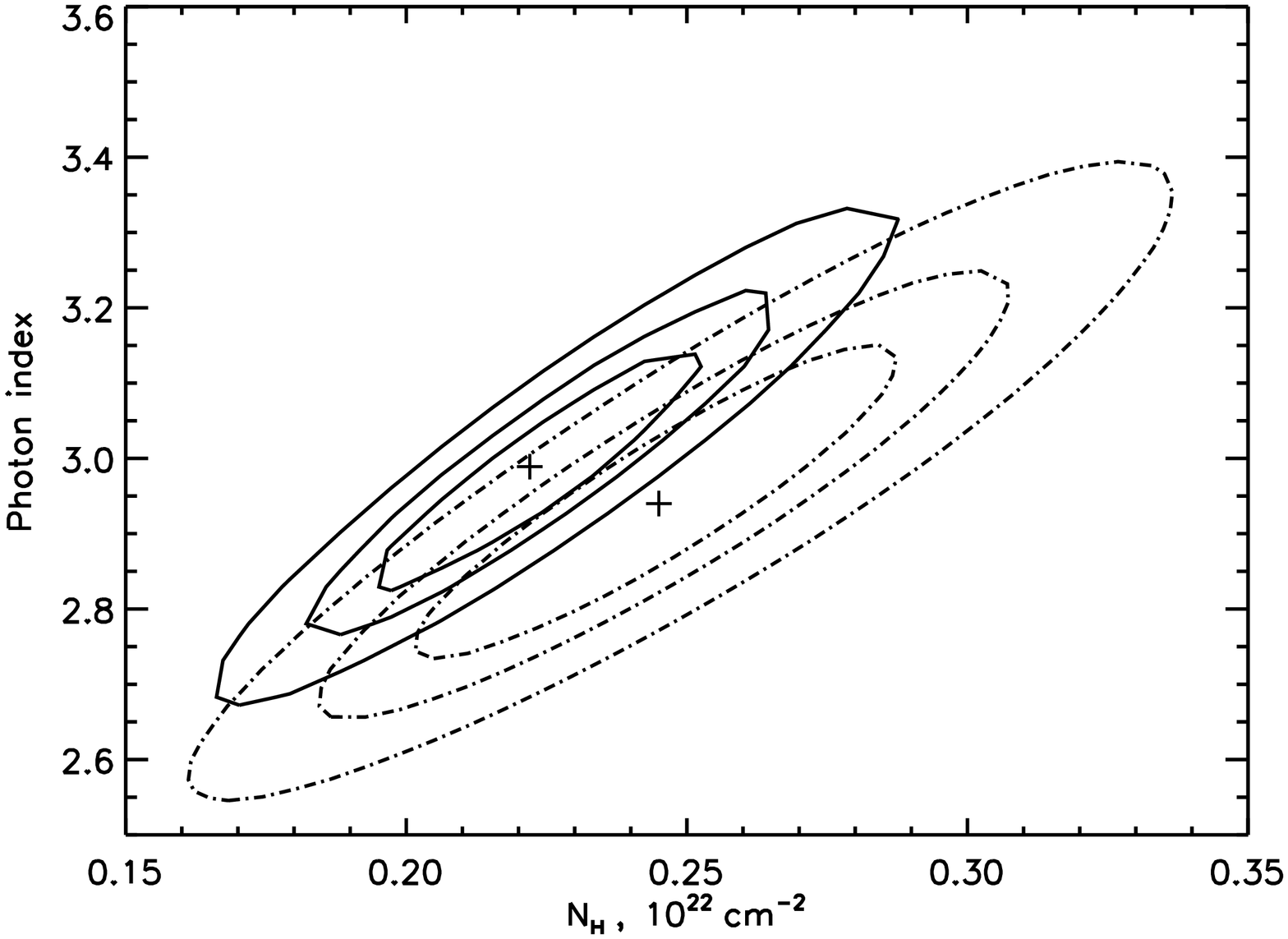}\\
\includegraphics[height=8cm,angle=90]{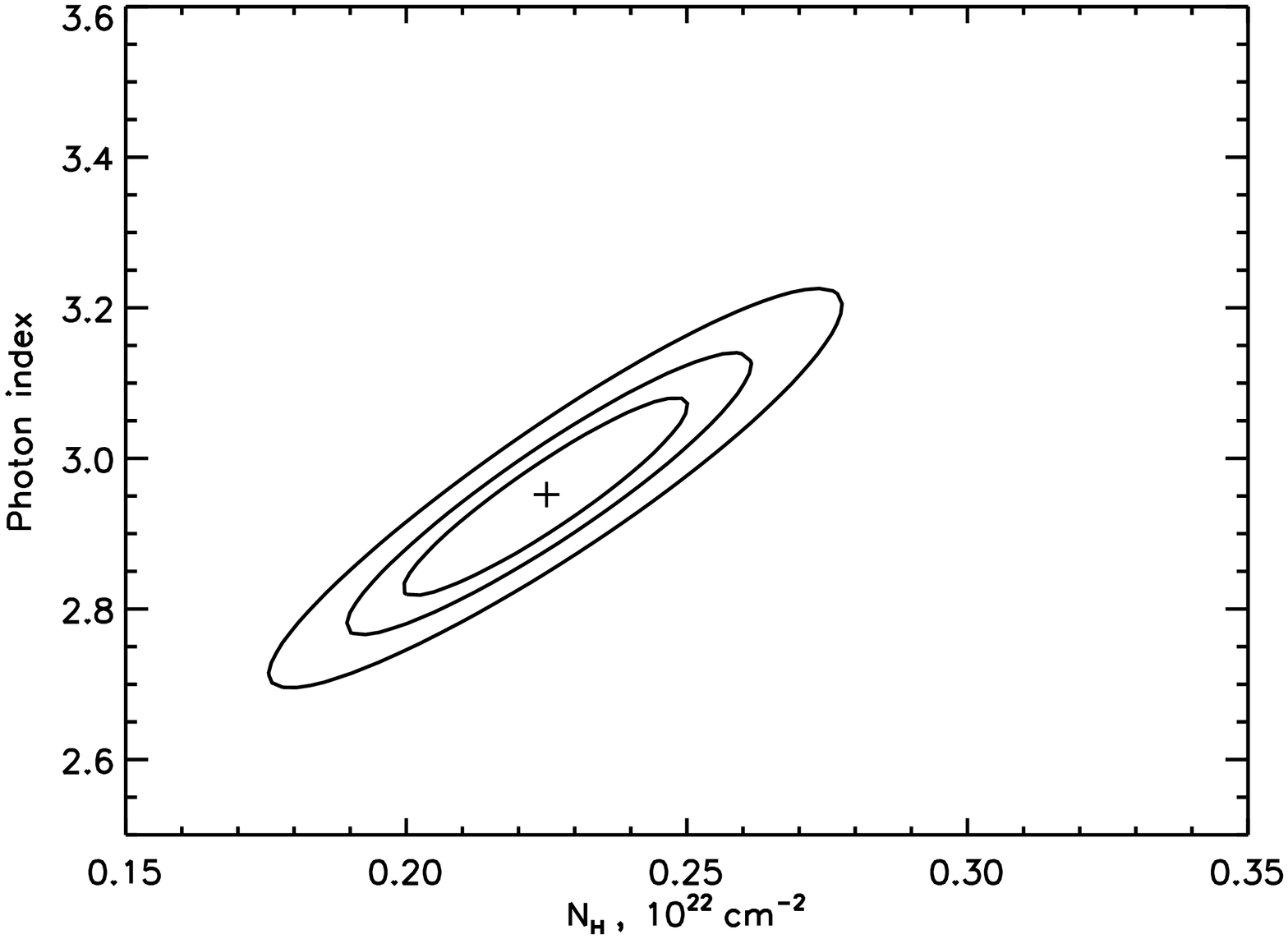}
\end{center}
\caption{{\em Top:} 68\%, 90\%, and 99\% confidence contours
for the {\sl Chandra} ACIS and  {\sl XMM-Newton} 
EPIC spectra of B1929 (dashed and solid lines, respectively), 
computed for two interesting parameters for the PL model.
 {\em Bottom:} 68\%, 90\%, and 99\% confidence contours for the
{\sl Chandra} and {\sl XMM-Newton} observations fitted simultaneously with
 the PL model.
\label{fig-individual-contours-pl} }
\end{figure}   
\clearpage

\clearpage
\begin{figure}
\begin{center}
\includegraphics[height=8cm,angle=90]{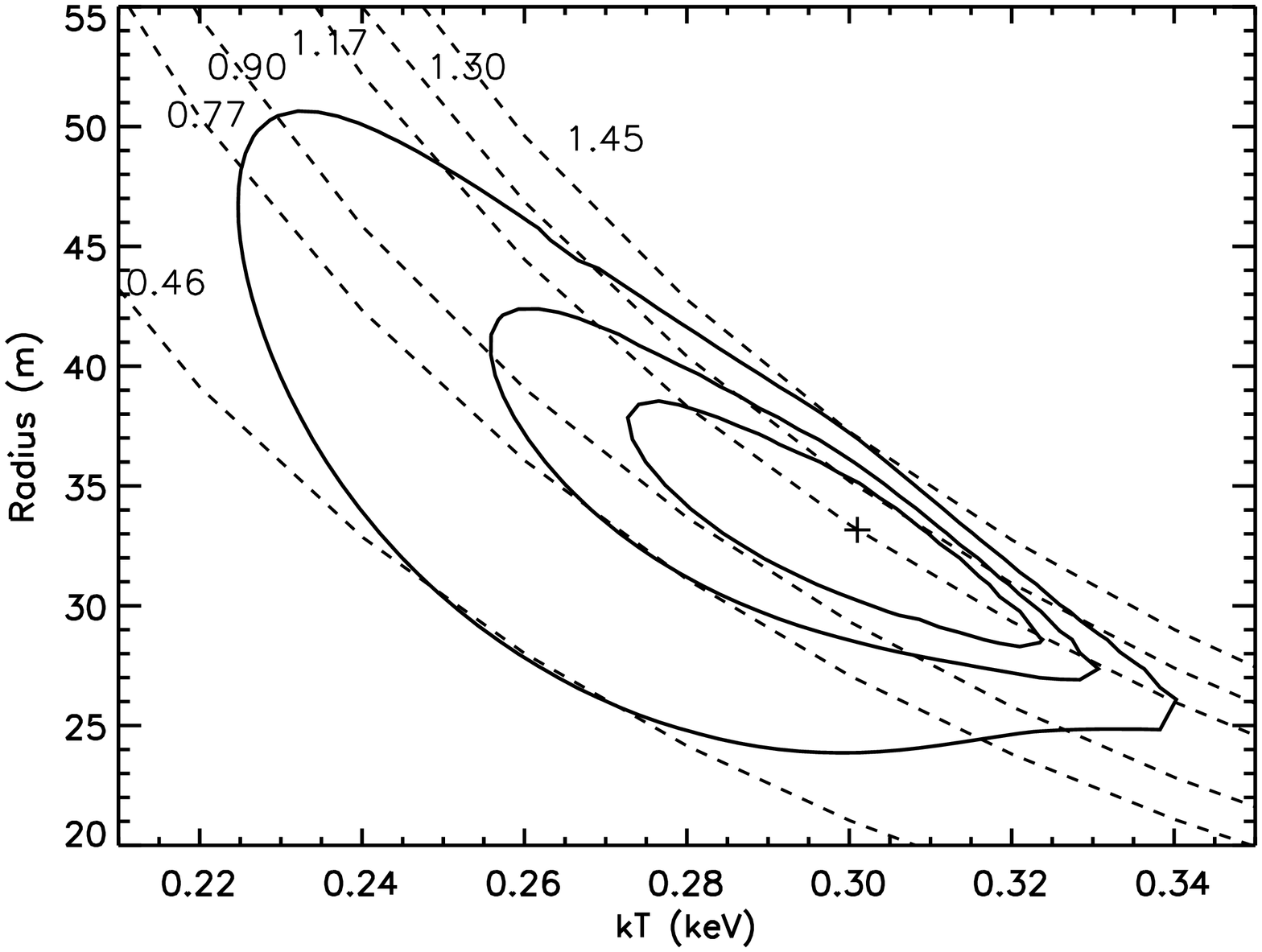}\\
\includegraphics[height=8cm,angle=90]{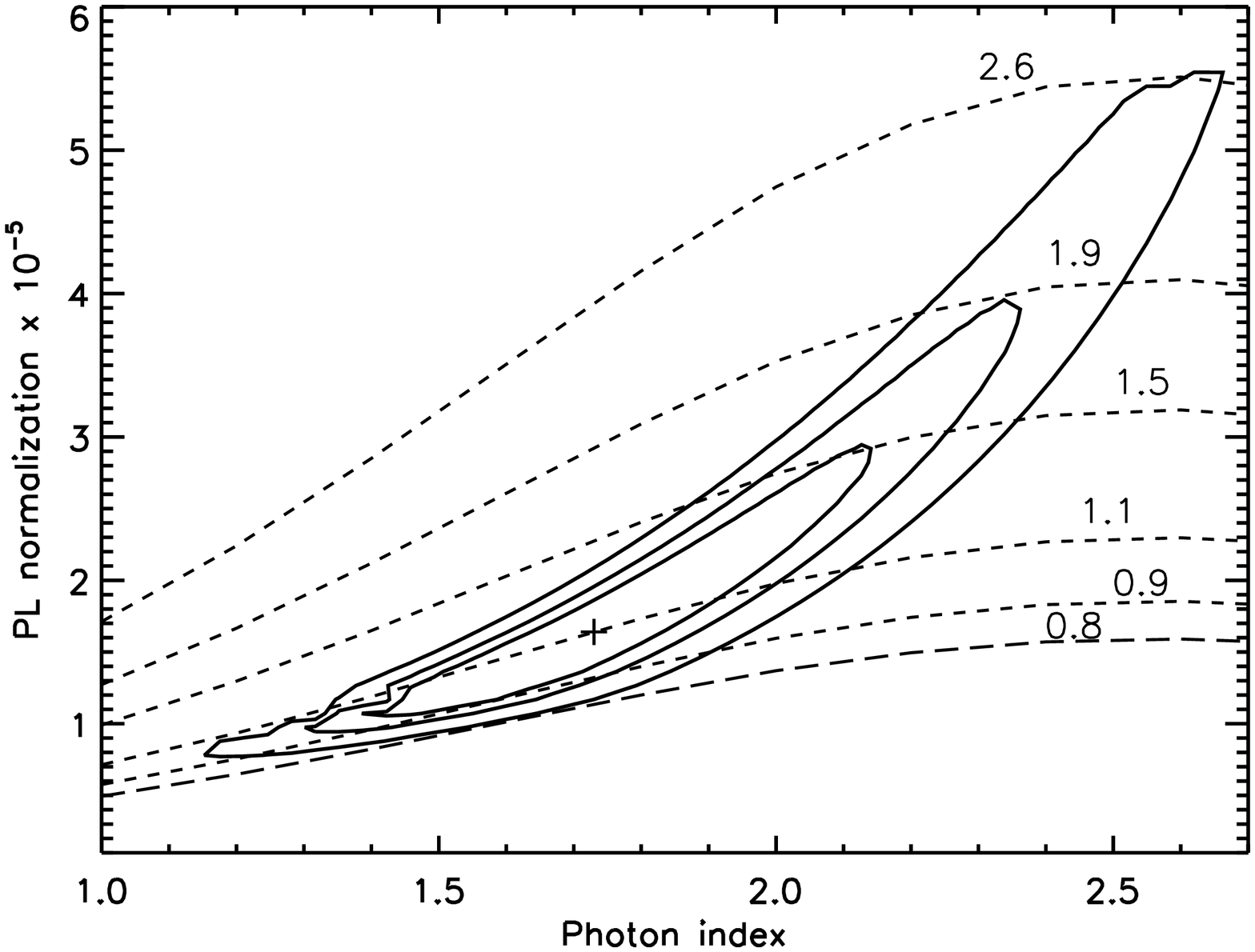}\\
\includegraphics[height=8cm,angle=90]{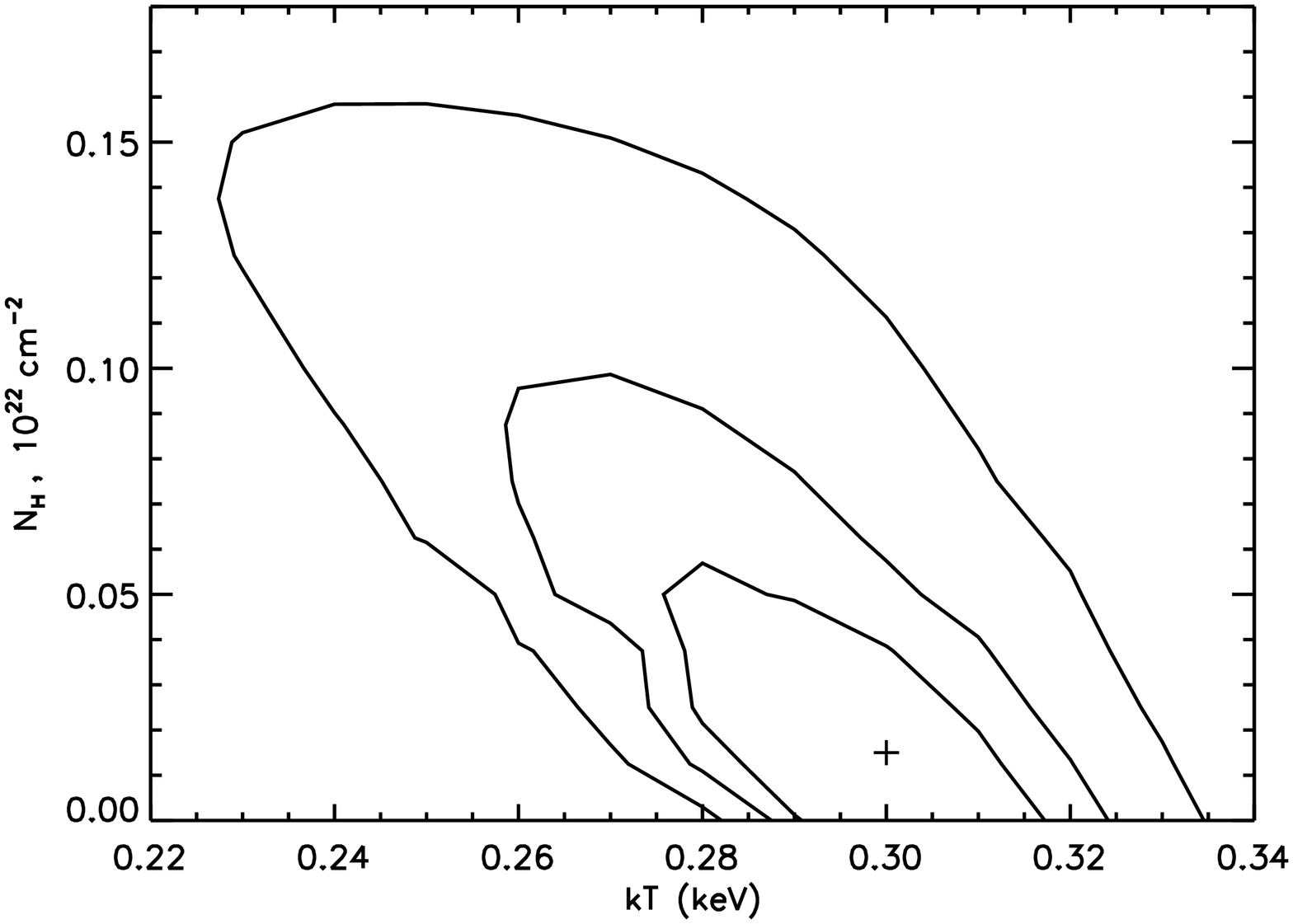}
\end{center}
\caption{68\%, 90\%, and 99\% confidence contours for various 
parameter pairs for the 
PL+BB fit to the spectrum of B1929. 
 The confidence levels correspond to two interesting parameters. 
The {\sl Chandra} ACIS-S and  {\sl XMM-Newton} spectra are fitted
 simultaneously 
with all parameters 
allowed to vary. The top panel shows the lines of constant bolometric
 luminosity
of an equivalent sphere (see \S\,3.2) in units of 
$10^{30}$\,ergs\,s$^{-1}$ for the BB component, 
while the middle panel shows the lines of constant unabsorbed flux 
in units of $10^{-13}$\,ergs\,cm$^{-2}$\,s$^{-1}$ for the PL
component. 
\label{cont-T-rad} }
\end{figure} 
\clearpage  

\subsection{Spectral analysis}
\label{spectral-analysis}

\subsubsection{The pulsar}
\label{spectral-analysis-pulsar}

The spectral {\sl Chandra} data reduction was done by applying the standard
 procedures
 in CIAO.
From a circular region with a radius
of $1.6''$ (or 3.3 pixels) centered on the pulsar, 
we extracted 645 and 690 counts 
from the first and second observation, respectively.
Four circular background regions, each with a radius of 5$\arcsec$\,,
  were selected  in the vicinity of
 the pulsar and then combined
and used for the background subtraction. 
The background contribution in the source aperture is negligibly small
($\sim 1$ count in each of the observations), but the aperture
may include a small number of 
photons emerging from the compact
 nebula. 
Using the radial profiles in
 Figure~\ref{fig-radial-profiles},
 we estimated the PWN contribution to be around 2\%--3\%;
hence it does not influence the results of the spectral fitting
significantly.
 We grouped the extracted counts in a minimum of 15 and 17 counts per energy
 bin for the spectral analysis.
The response matrices (rmf) were calculated
using the CIAO task {\sc mkacisrmf}, and the spectral analysis was
performed using XSPEC (ver.\ 11.3.2).

Since we did not find statistically significant changes between the 
two {\sl Chandra} observations, 
we fit both data sets simultaneously.
The absorbed power-law (PL) model yielded 
a very good
 fit  (Table~\ref{table-spectrum};
 Fig.\ \ref{fig-combined-spectra-pl}, top),
albeit with surprisingly large best-fit hydrogen column density,
$N_{\rm H}\approx2.5\times 10^{21}$ cm$^{-2}$, and photon index, 
$\Gamma\approx2.9$.
On the contrary, we were not able to produce a statistically
acceptable fit with the absorbed blackbody (BB) model  
as the discrepancy between the best-fit model and the data was large
below 0.7 keV and particularly above 2 keV, resulting
in $\chi_\nu^2 =1.86$ (vs.\ $\chi_\nu^2 =0.91$ for the PL fit)
for 71 d.o.f.

The pulsar models predict that a 
 thermal component from heated polar caps, and perhaps
from the whole NS surface, may be detectable,  making the main contribution
 at lower photon energies.  
 Therefore, we attempted to
fit the {\sl Chandra} spectra with the PL+BB model. 
The addition of the BB component, with the best-fit temperature
$kT\approx 0.3$ keV and emitting area $A\sim 4\times 10^3$ m$^2$,
resulted in a substantially smaller hydrogen column density,
 $N_{\rm H}<1.4\times 10^{21}$ 
cm$^{-2}$, and a softer PL spectrum, $\Gamma\approx 1.8$
(see Table~\ref{table-spectrum}),
but the probability that the BB component 
is required by the data was only 69.5\%, according to the F-test.
We also tested the  BB+BB model, but it yielded either  
statistically  unacceptable fits or  unphysical fitting parameters
(e.g., $kT\sim 100$\,keV, or an extremely low absorption column).

To compare  the {\sl Chandra} ACIS spectra with those
obtained with the
{\sl XMM-Newton} EPIC detector, we
extracted and fitted the pulsar spectra
from the three {\sl XMM-Newton}
observations. The source counts were extracted from a circular region with a 
radius of 13$''$ centered on the pulsar, 
while a 38$''$-radius circular region northwest
of the pulsar was used for the background subtraction.
 We collected a total of 450 (with 96.8\% of the source
contribution), 658 (94.8\%) and 805 (75.3\%) counts in the PN 
(small-window mode)
observations 0113051301, 0113051401, and 0113051501, respectively.
 In addition, we extracted
172 (98.0\%), 282 (97.3\%), and 110 counts from the MOS1, and 153 (97.3\%), 276
(98.0\%), and
 121 (96.8\%) from the MOS2 observations, with MOS cameras
operated in the large window mode. The EPIC response matrices and 
ancillary files
 were calculated using the SAS  tasks {\sc rmfgen} and
 {\sc arfgen}.
From the number of the PWN counts detected with ACIS in an aperture of $13''$,
we estimate that the contribution from the nebula in the EPIC spectra is 
about 5\%.

Fitting all the 9 {\sl XMM-Newton} data sets (PN, MOS1 and MOS2 in the three 
observations) simultaneously, 
we found that the PL model produced a statistically
 acceptable fit, with the model parameters consistent with those obtained 
from the {\sl Chandra} data 
(Fig.~\ref{fig-individual-contours-pl})\footnote{We should mention that 
our PL fitting parameters differ significantly from
those obtained by Becker et al.\ (2006) from fitting the
combined {\sl XMM-Newton} EPIC
and {\sl ROSAT}\, PSPC data (they do not provide model parameters from fitting
the {\sl XMM-Newton} data separately). 
For instance, these authors obtained a lower hydrogen column density,
 $N_{\rm H}=(1.6\pm 0.2)\times 10^{21}$ cm$^{-2}$, and 
a smaller photon index, $\Gamma = 2.72^{+0.12}_{-0.09}$. We believe that
the discrepancy is caused by systematic errors due to poor cross-calibration
of the PSPC and EPIC detectors.}. 
Adding the BB component to the spectral model resulted in about the same
fitting parameters as for the {\sl Chandra} data alone
(Table~\ref{table-spectrum}), but improved the fit
more significantly (94.8\%, according to the F-test).

Finally, we fit the {\sl Chandra} and {\sl XMM-Newton} data simultaneously,
first with the PL model.
Because of the 
larger PWN contribution 
in the {\sl XMM-Newton} data,
we first linked  all the model 
parameters except for
the PL normalization. However, since we found that the difference between the
PL normalizations 
in the {\sl Chandra} and
{\sl XMM-Newton} spectra was within the statistical errors,
we finally linked all the model parameters.
As expected, we obtained the PL parameters similar to those obtained from
the separate fits to the {\sl Chandra} and {\sl XMM-Newton} data, but
with smaller uncertainties (Fig.~\ref{fig-individual-contours-pl}, bottom).

Fitting the combined data with the PL+BB model, we found that the
addition of the BB component provided a very significant
improvement of the fit (99.98\%, according to the F-test).
The best PL+BB fit 
is shown in Figure~\ref{fig-combined-spectra-pl},
the fitting parameters are given in Table~\ref{table-spectrum},
and
the confindence contours for various model parameters are presented in 
Figure~\ref{cont-T-rad}.  We note that the hydrogen column density,
 $N_{\rm H}<4\times 10^{20}\,{\rm cm}^{-2}$, 
is significantly lower in the PL+BB fit and consistent with the value 
estimated from the pulsar's dispersion measure, 
while the slope of the PL components, $\Gamma\approx 1.7$, is similar to
those found in many other pulsars
(see \S\,\ref{spectral-discussion} for further discussion).

\clearpage
\begin{figure}
\begin{center}
\includegraphics[height=8cm,angle=90]{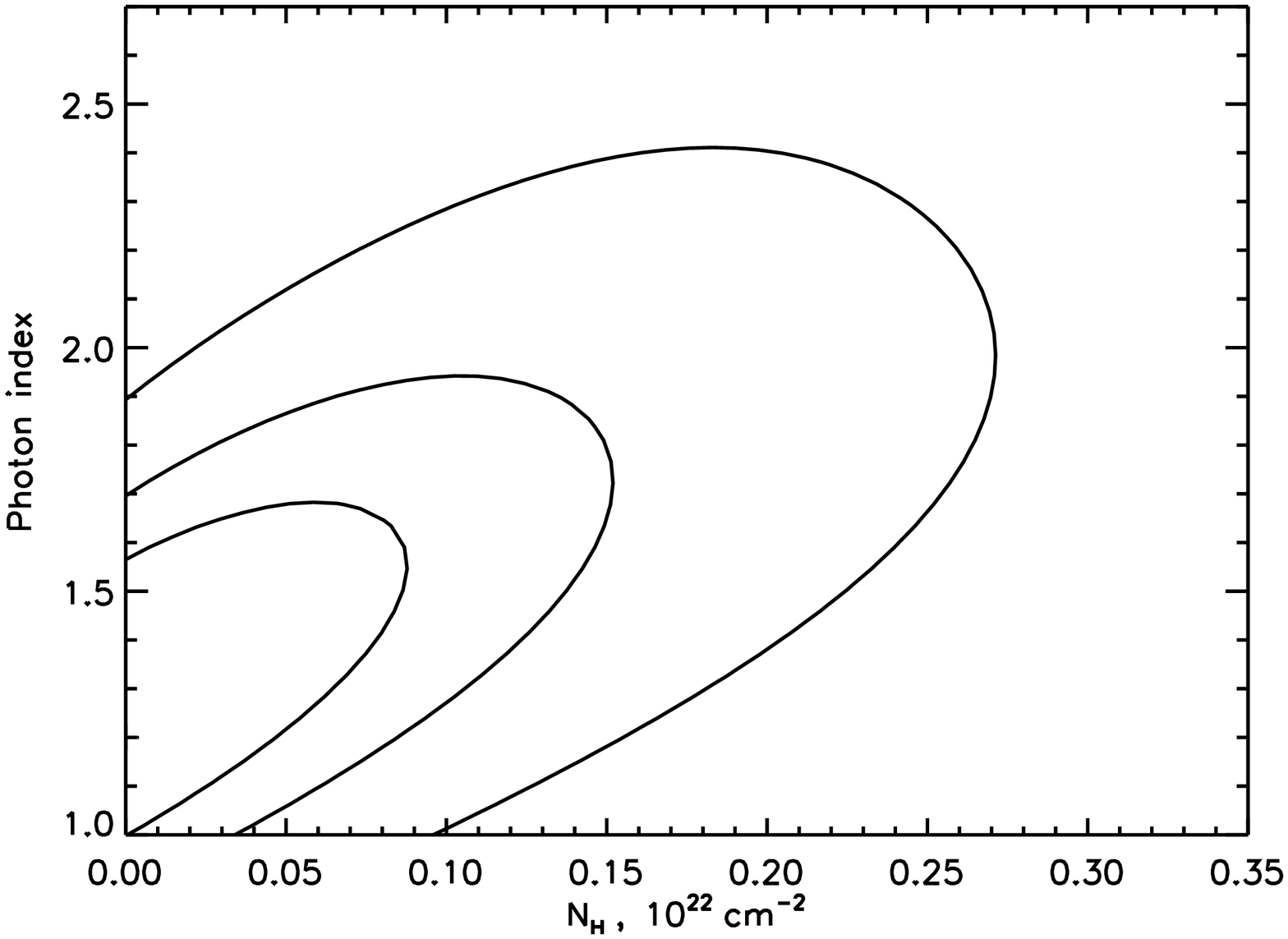}\\
\includegraphics[height=8cm,angle=90]{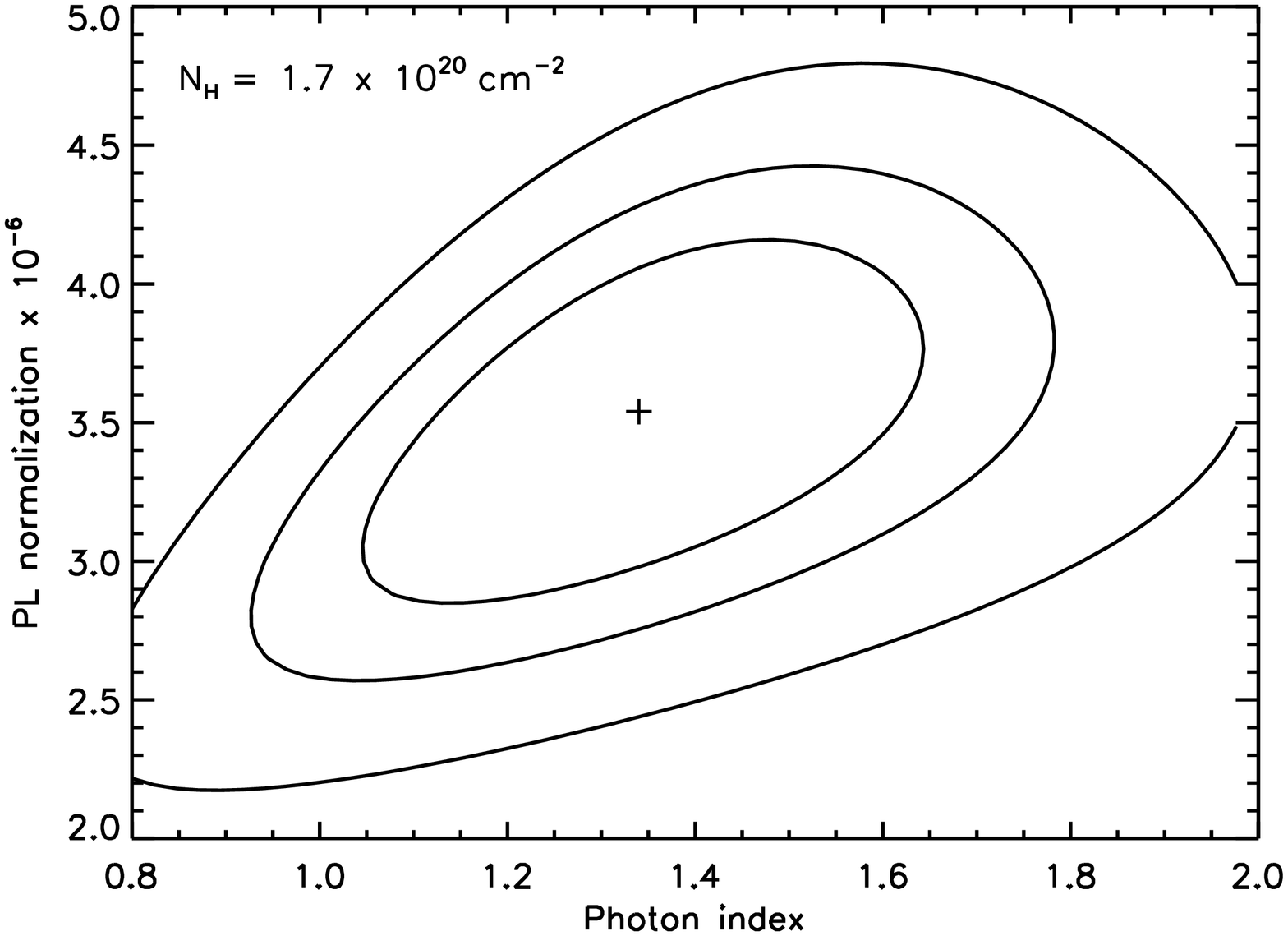}
\end{center}
\caption{{\em Top:} 
 68\%, 90\%, and 99\% confidence contours in 
the $N_{\rm H}-\Gamma$ plane
 of the  PL model  for the PWN spectral data 
(the elliptical region in Fig.~\ref{fig-combined-conv-smooth-image}).
{\em Bottom}: 68\%, 90\%, and 99\% confidence contours in 
the 
 $\Gamma$--Normalization plane
 of the PL model with the absorption
column fixed to the best-fit value obtained from the spectrum of the pulsar.
\label{cont-PWN}} 
\end{figure}   
\clearpage

\subsubsection{The nebula}
\label{spectrum-nebula} 

As the detected PWN is very faint, we have to choose a sufficiently large extraction region
to measure its spectrum.
The large-scale PWN
 appears to consist of  
a smaller, slightly brighter part up to 30$''$ behind the pulsar (marked by
the ellipse in the middle panel of 
Figure~\ref{fig-combined-conv-smooth-image}), 
and a fainter emission 
farther down, up to a distance of
 $\sim$2$'$.
 From the brighter part 
of the nebula region (the ellipse) we collected 138
and 136 counts (64\% and 63\% of which are from the nebula)
 in the first and second
 observation, respectively.

To constrain the model parameters, we fit  the spectra from the two observations simultaneously, 
with the absorbed PL model. 
The fit yielded a PL slope 
$\Gamma$=1.0--1.7  at a 90\% confidence level (Fig.\ \ref{cont-PWN}, top) 
and an absorption column 
 $N_{\rm H} = 0.0-0.9 \times  10^{21}$ cm$^{-2}$ (90\% confidence).
We also fitted the PWN spectra with the absorption column fixed to the
best-fit value of $N_{\rm H} = 1.7 \times  10^{21}$ cm$^{-2}$ obtained from
the spectrum of the pulsar, but the photon index range remains the same
(Fig.\ \ref{cont-PWN}, bottom).
We also attempted to fit the spectra from the PWN region farther down the
tail, to
see if there is any indication of spectral softening, but the number of
photons from this faint part of the nebula is too small for a miningfull
spectral analysis.   

The PWN flux, $F_{\rm PWN}=2.1$--$3.6\times10^{-14}$,
 determined from the 
elliptical region, is poorly constrained because of the low statistics.
We also measured a  flux of 
 $3.4-5.8\times10^{-14}$ ergs~cm$^{-2}$~s$^{-1}$ from the whole visible PWN.
The unabsorbed luminosity of the whole nebula estimated from this 
observed flux 
is 
$L_{\rm PWN} = 5.3$--$8.6\times 10^{29}$\,ergs\,s$^{-1}$, in the 0.3$-$8\,keV band.

The absorption column determined from the spectral fit of the nebula is
 significantly lower
than that determined from the 
PL fit of the pulsar spectrum, but it is in a good agreement with the value
derived from the PL+BB model. This further supports our detection of the thermal component in the spectrum of B1929.

\section{Discussion and conclusions}
\label{discussion}

\subsection{The nebula around PSR\,B1929+10}

\subsubsection{PWN morphology}

In the analysis of the two ACIS-S observations of B1929, we detected a faint 
PWN surrounding this old pulsar.
We found that the pulsar is
 immersed in a very compact, $\approx 9''\times 5''$ 
nebula
elongated in the
direction perpendicular to the pulsar's proper motion
(Fig.\ 1, bottom). 
Together with the two patchy ``wings'' (best seen in slightly smoothed
images), the overall compact structure 
looks like a bow shock PWN, with an apex at $\approx 3''$
($\approx 1.6\times 10^{16}$ cm at $d=361$ pc)
 ahead of 
the pulsar. 
In addition, there is marginal evidence of
a short jet-like feature emerging 
from the pulsar between the wings, in the direction opposite to 
the pulsar's proper motion  
(seen up to $3''$--$4''$ from the pulsar in the unsmoothed combined image 
and in the pulsar-subtracted image in Fig.\ \ref{fig-unsmoothed-images}).
 There is also some very
faint emission  in front of the 
apparent bow-shock apex, seen up to $6''$--$7''$ from the pulsar
(see Fig.\ \ref{fig-profile}).

In addition to the compact PWN, 
a faint, inhomogeneous
 tail-like structure was detected, extending
 in the direction approximately opposite to the pulsar's proper motion up to 
at least $\sim1.5'$--2$\arcmin$. This  
structure is well aligned with 
the much longer tail (up to 10\arcmin--15\arcmin)
detected in the {\sl ROSAT} and {\sl XMM-Newton} data 
\citep{1993Natur.364..127W,2006ApJ...645.1421B}.

The surface brightness of the detected 
PWN is not uniform: e.g., the two wings are
brighter than the other PWN structures
(see Table~\ref{table-regions}). 
Furthermore, the {\sl Chandra\/} and {\sl XMM-Newton\/} data suggest 
a possible variability of the detected PWN, although
the number of detected photons was too small to establish it unambiguously.

The approximate axial symmetry with respect to the direction
of the pulsar's proper motion strongly suggests that the PWN 
is shaped by the oncoming flow of the ambient matter in the pulsar's
reference frame. 
The corrected pulsar's proper motion and distance correspond to
the pulsar's speed $v_{\rm PSR} = 178\, (\sin i)^{-1}$
km s$^{-1}$, where $i$ is the 
angle between the velocity
vector and the line of sight.
It exceeds the ISM sound speed,
$c_s = (5kT/3\mu m_H)^{1/2} = 37 \mu^{-1/2} T_5^{1/2}$ km s$^{-1}$
(where $\mu$ is the molecular weight, $T_5=T/10^5$ K), for
$T< 2.3\times 10^6 \mu\, (\sin i)^{-2}$ K, i.e., for virtually any
plausible ISM temperature, which supports the bow-shock interpretation
of the detected nebular emission.

Our current understanding of the pulsar wind dynamics in a bow-shock
PWN mostly relies upon numerical models. A recent work in this
direction has been done by \citet{2003A&A...404..939V}, \citet{2005A&A...434..189B}, and \citet{2007MNRAS.374..793V}.
These models show that at supersonic pulsar speeds
the  termination shock (TS) of an {\em isotropic} pulsar wind
 acquires a bullet-like shape,
 with a distance 
\begin{equation}
R_h \simeq
(\dot{E}/4\pi c p_{\rm ram})^{1/2}\, 
\end{equation}
 between the pulsar and the
bullet head, where 
\begin{equation}
p_{\rm ram} = \rho_{\rm amb}\,v_{\rm PSR}^2
\end{equation}
 is
the ram pressure, 
and $\rho_{\rm amb}$ is the ambient density.
 The shocked pulsar wind is confined between the TS and
the contact discontinuity
 (CD) surface, while the forward bow
shock (FBS) separates the shocked ambient medium (between the CD and the
FBS) from
the unshocked one.
 For large Mach numbers, $\mathcal{M}\equiv v_{\rm \sc PSR}/c_{s}
=(3p_{\rm ram}/5p_{\rm amb})^{1/2} \gg 1$, and
small values of the magnetization parameter\footnote{The magnetization
parameter in the pre-shock wind is defined  
as the ratio of the Poynting flux to the kinetic energy flux.} 
of the
pre-shock pulsar wind, $\sigma < 0.1$,  the bullet's
cylindrical radius is $r_{\rm TS}\sim R_{h}$ 
 and the distance of
its back surface from the pulsar is $R_{b}\sim 6R_{h}$ 
\citep[see Figs.\ 1--3 in][]{2005A&A...434..189B}.
The shape of the CD surface ahead of the pulsar is similar to that of the
TS but with the apex
at $R_{\rm CD}\approx 1.3 R_h$, while  it acquires
a cylindrical shape with a radius $r_{\rm CD}\approx 4 R_h$ behind the
 TS bullet.
The shocked wind flows away from the pulsar within this cylinder,
 forming a PWN tail. The flow
velocity in the central part of the tail (the inner channel, 
$r\lesssim r_{\rm TS}$) is $\sim 0.1c$--$0.2c$,
while it is as high as $v_{\rm tail}=0.8c$--$0.9c$ in the bulk
 of the tail's volume,
at $r_{\rm TS} \lesssim r < r_{\rm CD}$ (where $r$ is the cylindrical
radius). The magnetic field, purely
toroidal in these models, also depends on $r$, being enhanced toward
the tail axis and the CD surface. Its typical value in the tail is
\begin{equation}
B_{\rm tail}\sim 2.8\, (c/v_{\rm tail})\, (\sigma p_{\rm ram})^{1/2} {\rm G},
\end{equation} 
where the $p_{\rm ram}$ is in units of dyn cm$^{-2}$
\citep[see eq.\ (13) of][]{2005A&A...434..189B}.
 In the numerical simulations, the flow velocity and the magnetic field
do not show substantial changes along the tail,
but these simulations are limited to relatively short
distances from the pulsar 
\citep[e.g., $<13 R_h$ in][]{2005A&A...434..189B}.

To interpret the B1929 PWN 
in the framework of 
the current MHD models, we attempted to match the observed
 morphological structures with those
predicted by the simulations. 
As the brightest X-ray emission is expected from the shocked
wind between the TS and CD in the front part of the PWN,
we assume that the very compact structure in the immediate vicinity
of the pulsar (Fig.~\ref{fig-unsmoothed-images}, bottom)
 is the PWN head and estimate the distance from the pulsar
to the CD apex to be $\sim 3''$.
This corresponds to
$R_{\rm CD}\sim 1.6\times 10^{16}$ cm,
$R_h \sim r_{\rm TS}\sim 1.2\times 10^{16}$ cm (or $2.3''$),
$R_b\sim 7.5\times 10^{16}$ cm (or $14''$), 
and $r_{\rm CD}\sim 4.8\times 10^{16}$ cm (or $9''$). 
In this picture, the patchy wings in the PWN image are produced by
the shocked wind between the TS and CD,  while the ``blobs'' farther
behind the pulsar are in the pulsar tail.
However, the faint emission ahead
of the pulsar (region 3, or the Front, in Fig.~\ref{fig-unsmoothed-images}, top, and Table~\ref{table-regions})
 is not explained in this model. We can only speculate that this emission might
be produced by a polar outflow along the pulsar's spin axis
(not included in the models), possibly scattered by the head wind,
 or it might suggest that, because of the 
Rayleigh-Taylor instability, the CD surface
acquires an irregular shape ahead of the pulsar, with ``fingers'' of the
shocked wind penetrating into the shocked ISM.
As the above estimates of sizes of the PWN elements are based on simplified
models and shallow images, one should consider them as order-of-magnitude
estimates rather than accurately established values. Therefore,
in the following estimates, we will
retain the explicit dependence on $R_h$, scaling it as 
$R_h=R_{h,16}\times 10^{16}$ cm.

Using this estimate for $R_h$ and equation (1), we can calculate the 
ram pressure:
\begin{equation}
p_{\rm ram}=
\dot{E}f_\Omega (4\pi cR_h^2)^{-1} = 
1.0\times 10^{-10} 
f_\Omega R_{h,16}^{-2}\,\,
{\rm dyn}\,\,{\rm cm}^{-2},
\end{equation}
where 
we introduced the factor $f_\Omega$ 
to account for possible anisotropy of the pulsar outflow
(e.g., $f_\Omega <1$ for a mostly equatorial outflow if the
equatorial plane is perpendicular
to the pulsar's velocity). Since, on the other hand,
\begin{equation}
p_{\rm ram}=\rho_{\rm amb}V_{\rm PSR}^2 =5.3\times 10^{-10} n_{\rm amb}\, (\sin i)^{-2}\,\,{\rm dyn}\, {\rm cm}^{-2}\,,
\end{equation}
where $n_{\rm amb}\equiv \rho_{\rm amb}/m_H$, we obtain an estimate
for the ambient number density: 
\begin{equation}
n_{\rm amb}= 0.19 f_\Omega R_{h,16}^{-2} \sin^2i\,\,\,  {\rm cm}^{-3}\,.
\end{equation}

Using equation (4), we can estimate the Mach number:
\begin{equation}
\mathcal{M}= 
7.7 f_\Omega^{1/2}R_{h,16}^{-1}p_{\rm amb,-12}^{-1/2}\,,
\end{equation}
where we scaled $p_{\rm amb}$ to $10^{-12}$ dyn cm$^{-2}$, a typical value
for the ISM.
From the corresponding sound speed, 
$c_s = 23 f_\Omega^{-1/2} R_{h,16} p_{\rm amb,-12}^{1/2} (\sin i)^{-1}$
km s$^{-1}$, we estimate the temperature:
\begin{equation}
T=3.9\times 10^4 \mu f_\Omega^{-1} R_{h,16}^2 p_{\rm amb,-12}(\sin i)^{-2}\, {\rm K}\,.
\end{equation}
The ambient pressure in the unperturbed ISM is in the range
$p_{\rm amb,-12}\approx 0.2$--2 
\citep{2001ApJ...551L.105H,2001RvMP...73.1031F}.
The pressure in the pulsar's neighborhood might be somewhat higher if
the pulsar's UV and soft X-ray emission heats and ionizes the surrounding
medium, which is supported by the lack of an H$_\alpha$
bow-shock PWN in the observations reported by \citet{2006ApJ...645.1421B}
\citep[see][for a detailed discussion on conditions
of observability of H$_\alpha$ bow-shock PWNe]{2001A&A...375.1032B}.
As the factor $\mu f_\Omega^{-1} R_{h,16}^2 (\sin i)^{-2}$ is also very likely
greater than unity, we expect $T\sim 10^5$ K to be a realistic estimate 
for the ambient temperature.

We should note that although the comparison of our data with the numerical
models leads to reasonable estimates for the parameters of the ambient
medium, these two-dimensional
 models are based on a number of assumptions that are not
necessarily realistic.  
In particular, the models assume an isotropic
pulsar outflow, whereas we know that it is predominantly equatorial, at
least in the case of young pulsars.
Obviously, a substantial anisotropy would distort the shape of the 
bow shock in the pulsar vicinity (e.g., decreasing $R_h$ in the case
of an equatorial outflow in the plane perpendicular to the pulsar
velocity).
For B1929, a hint of an equatorial outflow might be suggested by
the elongation of the most compact PWN structure 
perpendicular to the proper motion (Fig.~\ref{fig-unsmoothed-images}, bottom).
 Such a
structure, together with the ``wings'',
 might be interpreted as a torus of the shocked equatorial wind,
compressed and possibly bent by the oncoming wind.
Such an interpretation might also be supported by the marginal detection 
of the jet-like structure immediately behind the pulsar and the faint
emission ahead of the pulsar (a crushed counter-jet?).

Application of the PWN simulations to the observed tail
leads to some discrepancies. For instance, although the tail 
may look like 
a cylinder with the predicted diameter of $\sim 18''$ at the relatively
small distances, $\lesssim 2'$, from the pulsar in the ACIS images, 
its transverse
size apparently increases 
at larger distances, possibly reaching
$\sim 5'$ at the edge of the MOS field of view ($15'$ from the pulsar).
The tail's surface brightness looks very patchy, which might suggest
some interaction of the shocked wind with the surrounding medium,
such as the Kelvin-Helmholtz instability caused by the shear at the CD
surface. This instability
may produce clumps of shocked ISM embedded in the
shocked wind, which would distort the flow pattern and compress the
wind material \citep{2005A&A...434..189B}. 
The same instability may result in temporal variations in the X-ray
emission of the tail, with a time scale of 
$r_{\rm CD}/c \sim 0.5$ months, which might explain the alleged
variability of the PWN (\S~\ref{variability}).

Using equations (3) and (4), the magnetic field in the tail can be estimated
as
\begin{equation}
B_{\rm tail} = 
56 \sigma^{1/2} (0.5c/v_{\rm tail}) f_\Omega^{1/2}
R_{h,16}^{-1}\,\,\,\mu{\rm G}\,.
\end{equation}
We should note that this field becomes rather small, comparable to the interstellar magnetic field, at the  usually assumed small values of
the magnetization parameter, $\sigma\lesssim 10^{-2}$.

Using this estimate of $B_{\rm tail}$, we can estimate the synchrotron
cooling time for relativistic electrons
in the tail, $\tau_{\rm syn}=5.1\times 10^{10} \gamma_8^{-1}B_{-5}^{-2}$ s,
where $\gamma=10^8 \gamma_8$ is the electron Lorentz factor
and $B_{-5}=B/10^{-5}\,{\rm G}$. As the characteristic energy of synchrotron
photons is
\begin{equation}
E \sim 0.5 B_{-5} \gamma_8^2\,\,\, {\rm keV},
\end{equation}
photons with energy $E$
are emitted by electrons with $\gamma_8\sim 2 (E/1\,{\rm keV})^{1/2}
B_{-5}^{-1/2}$, 
and the synchrotron cooling
time corresponding to $E=8$ keV photons (upper energy of our band) is
\begin{equation}
\tau_{\rm syn}= 6.8\times 10^8 (\sigma f_\Omega)^{-3/4} 
(v_{\rm flow}/0.5c)^{3/2}
(E/8\,{\rm keV})^{-1/2} 
R_{h,16}^{3/2}\,\, {\rm s}.
\end{equation}
It follows from this equation that the projected tail length,
$l_{\rm tail}\sim v_{\rm flow} \tau_{\rm syn}\,\sin i$, as observed
at energy $E$, can be estimated as
\begin{equation}
l_{\rm tail} \sim 
1\times 10^{19}
\left(\frac{v_{\rm flow}}{0.5c}\right)^{5/2} 
\left(\frac{8\,{\rm keV}}{E}\right)^{1/2} 
\left(\frac{R_{h,16}^2}{\sigma f_\Omega}\right)^{3/4}\sin i\,\,\, {\rm cm}.
\end{equation}
For instance, for $\sigma=0.01$, $R_{h,16}f_\Omega^{-1/2}=2$, 
$v_{\rm flow}=0.5c$, $\sin i=0.5$, and $E<8$ keV
we obtain $l_{\rm tail}\gtrsim 150$ pc (i.e., $\sim 24^\circ$), 
two orders of magnitude larger than observed by
{\sl XMM-Newton} and {\sl ROSAT}. The main reason for this discrepancy
is the assumption that the flow speed remains very high along the 
entire tail in the
ideal MHD model. \citet{2005A&A...434..189B}
have noticed that a similar discrepancy arises
when the same model is applied to the 
X-ray tail of the Mouse PWN: the observed
tail length is much shorter than that estimated from the model at
any $\sigma <1$. These authors suggest that the 
flow can be slowed down by the interaction with the ambient
medium,
which could also explain the 
divergence 
 of the Mouse tail in the radio.
As these factors have not been included in the current models,
we can only empirically estimate an average flow velocity that would
be consistent with the observed length of 1.5 pc:
$v_{\rm flow} \sim 0.1 c\, \sigma_{-2}^{3/10} 
(f_\Omega/R_{h,16}^2)^{3/10} (\sin i)^{-2/5}$,
assuming that equation (3) is still applicable
(here $\sigma_{-2}=\sigma/10^{-2}$).
We note that this velocity is still much larger than the pulsar velocity.
This means that the 
equation 
\begin{equation}
l_{\rm tail}= v_{\rm PSR,\perp} \tau_{\rm syn}
\approx 0.3 (v_{\rm PSR,\perp}/178\,\,{\rm km}\,\,{\rm s}^{-1})
 \gamma_8^{-1} B_{-5}^{-2}\,\,\, {\rm pc},
\end{equation}
 often
used for estimating the magnetic field \citep[e.g.,][]{2006ApJ...645.1421B},
is inapplicable in this case\footnote{Equation (13) implies that the
shocked wind flow acquires the speed of the ambient ISM matter
in the immediate vicinity of the pulsar, so that the observed elongated PWN
is actually a ``trail'' of the decelerated wind left behind the moving pulsar.
Such an assumption strongly contradicts all the models of bow-shock PWNe
\citep[e.g.,][]{2005ApJ...630.1020R,2005A&A...434..189B}.
 \citet{2006ApJ...645.1421B} obtained
a reasonable estimate for the magnetic field, $B\lesssim 12$ $\mu$G,
 using eq.\ (13) because they assumed $\gamma=10^6$. However, at such
$B$ and $\gamma$,
the energy of a synchrotron photon is $E=6$ eV (see eq.\ [10]), well
below the X-ray band.}.

It is worthwhile to mention that 
the length of the tail of B1929 measured so far might be limited by the
  field of view of the available {\sl XMM-Newton} observations,
 and further deep observations along the tail are needed to determine its full extent.
Recent observations with {\sl Chandra} and {\sl XMM-Newton} have discovered
a number of other PWNe with very long tails \citep{2008AIPC..983..171K},
including the 6\,pc tail ($l_{\rm tail}/R_{h}\sim 600$) behind 
PSR J1509$-$5850, the longest pulsar tail known to date \citep{2008arXiv0802.2963K}.
Detection of more such objects
 would help to facilitate further modeling
  that would account for cooling in the varying magnetic field  at larger distances from the pulsar.

\subsubsection{X-ray luminosity and spectrum of the PWN}

From our  estimate of the total unabsorbed flux of the 
B1929 PWN detected with {\sl Chandra},
we 
estimated the X-ray PWN luminosity, $L_{\rm PWN}=(5.3$--$8.6)\times 10^{29}$
ergs s$^{-1}$, which corresponds to the PWN efficiency
 $\eta_{\rm PWN} \equiv L_{\rm PWN}/\dot E = 
(1.4$--$2.2)\times 10^{-4}$ in the 0.3--8\,keV band.
Similar values of the luminosity and efficiency have been measured
in the {\sl XMM-Newton} data
by \citet{2006ApJ...645.1421B} from a larger area farther
southwest
from the pulsar.
As both measurements were taken for small fractions of the PWN,
the luminosity and efficiency of the entire PWN can be higher than these
estimates.

The estimated efficiency of the B1929 PWN is within the range
$\eta_{\rm PWN}\sim 10^{-4.5}$--$10^{-2}$ found by \citet{2007ApJ...660.1413K}
in a recent study of several energetic middle-aged Vela-like pulsars and
 their PWNe. 
Compared to a few known PWNe with long tails, which, on average, show higher
efficiencies, $\eta_{\rm PWN}\sim 10^{-3.8}$--$10^{-1.7}$,
 than more compact PWNe \citep[see][]{2008arXiv0802.2963K}, 
the B1929 PWN is among the least efficient ones;
however, we stress that a deeper observation of B1929 is needed to 
measure the efficiency more accurately.

The spectral slope of the PWN spectrum is also consistent with the values  
measured for other PWNe \citep[e.g., see][]{2007ApJ...660.1413K,2008AIPC..983..171K}.

\subsection{Spectral properties of PSR\,B1929+10}
\label{spectral-discussion}

As we mention in \S\ref{intro}, it has been a matter of debate
 whether the X-ray 
emission from B1929 is predominantly magnetospheric or it has a significant
thermal component emitted from the neutron star surface.
Most recently, \citet{2006ApJ...645.1421B} have concluded, 
based on the analysis of
the {\sl XMM-Newton} data, that the spectrum of B1929 is best
described by a PL model (i.e.\  the emission is
predominantly magnetospheric),
while the contribution of the thermal component, modeled as BB radiation,
does not exceed $\sim$7\%. Our analysis of the {\sl Chandra} and
 {\sl XMM-Newton}
spectra has shown, however, that 
adding the BB component to the PL model significantly improves the fit to the
combined {\sl Chandra} plus {\sl XMM-Newton} spectrum, with the 
the best-fit BB component
providing $\sim$45\% of the luminosity in the 0.3--10 keV band.

In addition to the better fit, there are other serious arguments in favor of
 the
PL+BB model. First of all,  
the hydrogen column density, $N_{\rm H}=(2.23\pm0.27)\times 10^{21}$ cm$^{-2}$,
obtained from the PL fit, looks unreasonably large. In particular,
it is 
much larger than the standard estimate, 
$N_{\rm H}\sim 10 N_e \approx 1\times 10^{20}$ cm$^{-2}$, obtained
from the pulsar's dispersion measure, DM = 3.178 pc cm$^{-3}$, under
the usual assumption of a 10\% ISM ionization (where
$N_e$ is the electron column density). In other words, the PL model
requires a very low ISM ionization, $\sim$0.4\%, (i.e., a very large
ratio $N_{\rm H}/N_e \sim 230$, much larger than for any other
radio pulsar detected in X-rays). Also, taking into account that B1929 is
a nearby pulsar ($d=361$ pc),
the $N_{\rm H}$ obtained from the PL fit
is uncomfortably close to the HI column density, 
$N_{\rm HI}=3.5\times 10^{21}$, throughout
the entire Galaxy in the direction of B1929 \citep{2005A&A...440..775K}.
On the contrary,
$N_{\rm H}= 0.17^{+0.23}_{-0.17}\times 10^{21}$ cm$^{-2}$ inferred from the
PL+BB fit is consistent with the usual estimate based on the dispersion
measure as well as with the $N_{\rm H}$ estimated from the PWN spectrum
(see Fig.\ \ref{cont-PWN}),
 and it is much lower than the total Galactic $N_{\rm HI}$.

Also, the PL+BB model looks more attractive than the PL model
 because it gives the slope of the PL component,
$\Gamma\approx 1.7$, within the range of spectral slopes found 
for a large sample of young
pulsars \citep[$\Gamma \approx 1-2$;][]{2003ApJ...591..361G},
while the PL fit to the B1929 spectrum results in
a considerably steeper slope, $\Gamma \approx 3$.
One could argue that 
the PL fits to the spectra of other old pulsars \citep[e.g., 
B0950+08, B2224+65, B0823+26, B0628--28, 
B1133+16, B0943+10;][]{2004ApJ...616..452Z,2004ApJ...615..908B,2005ApJ...630L..57T,2006ApJ...636..406K,2005ApJ...624L.109Z}
also show rather steep slopes, $\Gamma\approx 2-3$, suggesting that
pulsar spectra might soften with increasing age or decreasing spin-down
power \citep{2006ApJ...636..406K}. However, 
a plausible alternative 
interpretation
of the softer spectra of old pulsars is
that they are, in fact, comprised of
a (soft) thermal component and a PL component with a more gradual 
slope\footnote{Note
that 
the observed spectra of these pulsars can be satisfactorily fitted
with quite different models
because 
these objects are very faint in the X-ray range.},
similar to those found in young pulsars \citep{2004ApJ...616..452Z}.

Thus, based on the goodness of fit and the astrophysical arguments,
we conclude that the X-ray emission from B1929 most likely includes both
 thermal 
and magnetospheric components.
The BB model for the thermal component gives the 
apparent temperature $kT_a\approx0.3$\,keV
and projected emitting area $A_{\perp,a}\sim 3000$ m$^2$, 
at the distance of 361 pc,
suggesting that the thermal emission originates from small heated spots
(e.g., polar caps). These temperature and area correspond to the
apparent radius $R_a^{\rm es}=(A_{\perp,a}/\pi)^{1/2}\approx 33$ m 
and bolometric luminosity $L_{{\rm bol},a}^{\rm es}=4 A_{\perp,a}\sigma T_a^4
\approx 1.1\times 10^{30}$ ergs s$^{-1}$ of an equivalent sphere.
The true size and luminosity of the polar caps depend on the geometry
and the gravitational redshift factor, $g_r=(1-R_g/R_{\rm NS})^{1/2}$,
where $R_g=2.95 M_{\rm NS}/M_\odot$ km, $R_{\rm NS}$ and $M_{\rm NS}$ are
the mass and radius of the neutron star. 
 If there are two identical hot spots at the poles of a centered magnetic
dipole, the polar cap radius 
and the luminosity of two
polar caps, as measured at the neutron star surface, are
$R_{\rm pc}=
R_{a}^{\rm es}f^{-1/2}$ and $L_{\rm bol,pc}=
L_{{\rm bol},a}^{\rm es}/(2fg_r^4)$, where $f\leq 1$ is a geometrical factor
depending on the angles $\zeta$ beween the line of sight and the
spin axis and $\alpha$ between the spin and magnetic axes, as well as
on $g_r$ \citep{2007ApJ...664.1072P}.
For instance, assuming $M_{\rm NS}=1.4 M_\odot$, $R_{\rm NS}=10$ km 
(i.e.\ $g_r=0.766$),
$\alpha=36.0^\circ$, and $\zeta=61.5^\circ$ \citep{2001ApJ...553..341E} and
using the approach described by \citet{1995AstL...21..149Z} and
  \citet{2002ApJ...566L..85B},
we find 
$f=0.897$,
which gives
 $R_{\rm pc}= 
1.06 R_{a}^{\rm es}\approx 35$ m and 
$L_{\rm bol,pc}=1.62 L_{{\rm bol},a}^{\rm es}
\approx 1.8\times 10^{30}\, {\rm ergs\, s}^{-1}=4.7\times 10^{-4}\dot{E}$.

The estimated polar cap radius is a factor of $\sim$9 smaller than 
the conventional polar cap radius of a radio pulsar,
$\tilde{R}_{\rm pc} = (2 \pi R_{\rm NS}^3/cP)^{1/2}$
\citep[see e.g.,][]{1991tnsm.book.....M}, which is
$\approx 300$ m for B1929, assuming $R_{\rm NS} =10$ km.
However, the observed 
polar cap radii can be substantially smaller
or larger than the conventional value.
In particular,
BB fits of several old pulsar spectra show $R_{\rm pc}\sim (0.1$--$0.2)\,
\tilde{R}_{\rm pc}$ 
\citep[e.g.,][]{2004ApJ...616..452Z,2005ApJ...624L.109Z,2006ApJ...636..406K}.
For some pulsars, 
such a discrepancy can be alleviated assuming that the polar cap
is covered by a hydrogen or helium atmosphere, in which case the effective
temperature would be a factor of 2 lower, and the radius a factor of 3--10
larger 
 while the bolometric luminosity would not change significantly 
\citep{1995lns..conf...71P,2004ApJ...616..452Z}. 
Another explanation for such a discrepancy
was proposed by 
\citet{2005ApJ...624L.109Z}, who suggested that only a small fraction of the polar cap surface,
associated with footprints of sparks produced by intermittent breakdowns of an ``inner gap'' above the polar cap, is hot enough to emit X-rays.

The luminosity of the detected thermal component, which is less dependent on
the assumed spectral model than the radius and temperature,
can be compared with the models for 
pulsar polar cap heating \citep{2002ApJ...568..862H,2001ApJ...556..987H}. 
For B1929, these models predict
the polar cap luminosity 
$L_{\rm bol,pc}\sim 
10^{31}$\,ergs\,s$^{-1}$ if the polar cap
 is heated by positrons produced through curvature radiation 
of electrons accelerated in the neutron star magnetosphere.
Although the polar cap thermal luminosity estimated from the
PL+BB spectral model is a factor
of a few lower than that predicted by \citet{2001ApJ...556..987H},
this can be considered as a reasonably good agreement, with account
of the uncertainties of both the theoretical model
 and the observational results.
We expect that deeper observations of other old pulsars would also detect
thermal components from their polar caps and help us better understand
the polar cap heating mechanisms.

As Figure~\ref{fig-combined-spectra-pl}  shows, the thermal component
 dominates at 
$0.5\,{\rm keV}\lesssim E \lesssim 2\,{\rm keV}$. This means that we can
expect different pulse shapes within and outside this energy range,
because the pulsations of thermal radiation should be smoother and
shallower than those of the magnetospheric radiation. 
The study of the energy dependence of pulsations not only can
confirm the presence of the thermal component, but it would also allow
one to infer the polar cap geometry and emission mechanism
(e.g., BB vs.\ a light-element atmosphere), and even measure the mass-to-radius
ratio for the neutron star \citep{1995AstL...21..149Z,1997ApJ...490L..91P,2004ApJ...616..452Z}.
 Since the current {\sl XMM-Newton} data with
sufficient time resolution do not
have enough counts for such an analysis, a new deep observation of
B1929 is needed to perform this important test.

The spectral slope
 of the PL (magnetospheric) component inferred from the PL+BB fit 
is $\Gamma\approx 1.7$,
similar to those of younger pulsars. The luminosity of this component,
 $L_X=
4\pi d^2 F_X^{\rm unabs} \approx 1.7\times 10^{30}$ ergs s$^{-1}$ in
the 0.3--10 keV band,
correponds to the nonthermal X-ray efficiency
 $\eta_{\rm nonth}=L_X/\dot{E}\approx 4.4\times 10^{-4}$, typical for
the whole population of radio pulsars detected in the X-rays, and comparable
to the thermal (polar cap) efficiency. Notice that the one-component PL fit
results not only in a much softer spectrum ($\Gamma\approx 3.0$) but also
in a higher luminosity, $L_X\approx 6.6\times 10^{30}$ ergs s$^{-1}$,
and efficiency, $\eta_{\rm nonth}\approx 1.7\times 10^{-3}$, above the
typical value for X-ray emitting radio pulsars. A similar trend
has been observed in other old pulsars: PL fits of their spectra yield
X-ray efficencies noticeably higher that those of young and middle-aged
pulsars, which might suggest that the X-ray efficiency in this energy
range grows with increasing age and decreasing spin-down power, perhaps
at the expence of the efficiency at higher photon energies.
However, this conclusion is based on the assumption that the contribution
of the thermal component is negligible in the soft X-ray range, which
we believe is not valid at least for the best-studied old pulsars,
B1929 and B0950+08. To understand the evolution of X-ray properties
of pulsars, deep observations of a larger sample of old pulsars are
warranted.

\subsection{Summary}

Two ACIS-S observations of B1929 revealed a faint 
PWN surrounding
 this old, nearby pulsar.
 The observed morphology includes 
 a compact nebula with two patchy wings in the immediate vicinity of B1929, and a tail
 extending in the direction opposite to the pulsar's proper motion, 
aligned with the much longer tail detected in the previous {\sl ROSAT} and
{\sl XMM-Newton}
 observations.  
The shape of the nebula and its spectral properties
 are  consistent with the proposed
bow-shock classification.
The properties of the compact nebula suggest that the pulsar wind outflow
is anisotropic, possibly concentrated toward the equatorial plane
perpendicular to the pulsar velocity. The size of the observed tail
implies an average flow velocity of $\sim 0.1 c$. The whole observed
PWN radiates about $2\times 10^{-4}$ of the pulsar spin-down power
in X-rays.
 By comparing the observed PWN properties with the bow-shock MHD models,
we estimated the temperature  of the local ISM to be $\sim10^5$\,K,
 which is consistent with the lack of
 the H$\alpha$ bow-shock emission around this pulsar.

In the spectral analysis of the combined {\sl Chandra} 
and {\sl XMM-Newton} data, we detected  a thermal component
in the pulsar radiation,
whose luminosity, $\sim (1$--$2) \times 10^{30}$\,ergs\,s$^{-1}$,
 is comparable with the magnetospheric  
luminosity in the X-ray band. 
The thermal radiation is likely emanating from polar caps
heated by positrons created by the 
curvature radiation of ultrarelativistic electrons accelerated in the
pulsar magnetosphere.
The spectrum and X-ray efficiency of the magnetospheric component are
similar to those found in young and middle-aged pulsars.
Further X-ray observations of the pulsar will allow one to
better characterize its spectral and timing properties and understand
the nature of the magnetospheric radiation and the mechanisms of polar
cap heating.

We thank Patric Broos for the useful discussion about the MARX  
simulations, Oleg Kargaltsev and Leisa Townsley for  the help
 with the high-resolution imaging techniques, and Slava Zavlin
for the discussion about the {\sl XMM-Newton} spectral results.
We wouls also like to thank the referee for very helpful comments
and suggestions.
This work was partially supported by NASA grant NAG5-10865
and {\sl Chandra} award SV4-74018.

\bibliographystyle{apj}
\bibliography{./paper}

\begin{thebibliography}{}

\bibitem[\protect\citeauthoryear{{Becker} et~al.}{{Becker}
  et~al.}{2005}]{2005ApJ...633..367B}
{Becker}, W., {Jessner}, A., {Kramer}, M., {Testa}, V.,  \& {Howaldt}, C. 2005,
  \apj, 633, 367

\bibitem[\protect\citeauthoryear{{Becker} et~al.}{{Becker}
  et~al.}{2006}]{2006ApJ...645.1421B}
{Becker}, W., et~al. 2006, \apj, 645, 1421

\bibitem[\protect\citeauthoryear{{Becker} \& {Tr\"umper}}{{Becker} \&
  {Tr\"umper}}{1997}]{1997A&A...326..682B}
{Becker}, W.,  \& {Tr\"umper}, J. 1997, \aap, 326, 682

\bibitem[\protect\citeauthoryear{{Becker} et~al.}{{Becker}
  et~al.}{2004}]{2004ApJ...615..908B}
{Becker}, W., {Weisskopf}, M.~C., {Tennant}, A.~F., {Jessner}, A., {Dyks}, J.,
  {Harding}, A.~K.,  \& {Zhang}, S.~N. 2004, \apj, 615, 908

\bibitem[\protect\citeauthoryear{{Beloborodov}}{{Beloborodov}}{2002}]{2002ApJ.%
..566L..85B}
{Beloborodov}, A.~M. 2002, \apjl, 566, L85

\bibitem[\protect\citeauthoryear{{Binney} \& {Tremaine}}{{Binney} \&
  {Tremaine}}{1987}]{1987gady.book.....B}
{Binney}, J.,  \& {Tremaine}, S. 1987, {Galactic dynamics} (Princeton, NJ,
  Princeton University Press)

\bibitem[\protect\citeauthoryear{{Boesgaard} \& {Tripicco}}{{Boesgaard} \&
  {Tripicco}}{1986}]{1986ApJ...303..724B}
{Boesgaard}, A.~M.,  \& {Tripicco}, M.~J. 1986, \apj, 303, 724

\bibitem[\protect\citeauthoryear{{Bucciantini}, {Amato}, \& {Del
  Zanna}}{{Bucciantini} et~al.}{2005}]{2005A&A...434..189B}
{Bucciantini}, N., {Amato}, E.,  \& {Del Zanna}, L. 2005, \aap, 434, 189

\bibitem[\protect\citeauthoryear{{Bucciantini} \& {Bandiera}}{{Bucciantini} \&
  {Bandiera}}{2001}]{2001A&A...375.1032B}
{Bucciantini}, N.,  \& {Bandiera}, R. 2001, \aap, 375, 1032

\bibitem[\protect\citeauthoryear{{Caraveo} et~al.}{{Caraveo}
  et~al.}{2004}]{2004MmSAI..75..470C}
{Caraveo}, P.~A., {Bignami}, G.~F., {De Luca}, A., {Pellizzoni}, A.,
  {Mereghetti}, S., {Mignani}, R.~P., {Tur}, A.,  \& {Becker}, W. 2004, Memorie
  della Societa Astronomica Italiana, 75, 470

\bibitem[\protect\citeauthoryear{{Chatterjee} et~al.}{{Chatterjee}
  et~al.}{2004}]{2004ApJ...604..339C}
{Chatterjee}, S., {Cordes}, J.~M., {Vlemmings}, W.~H.~T., {Arzoumanian}, Z.,
  {Goss}, W.~M.,  \& {Lazio}, T.~J.~W. 2004, \apj, 604, 339

\bibitem[\protect\citeauthoryear{{Cutri} et~al.}{{Cutri}
  et~al.}{2003}]{2003tmc..book.....C}
{Cutri}, R.~M., et~al. 2003, {2MASS All Sky Catalog of point sources.} (The
  IRSA 2MASS All-Sky Point Source Catalog, NASA/IPAC Infrared Science
  Archive.~http://irsa.ipac.caltech.edu/applications/Gator/)

\bibitem[\protect\citeauthoryear{{Del Zanna}, {Amato}, \& {Bucciantini}}{{Del
  Zanna} et~al.}{2004}]{2004A&A...421.1063D}
{Del Zanna}, L., {Amato}, E.,  \& {Bucciantini}, N. 2004, \aap, 421, 1063

\bibitem[\protect\citeauthoryear{{Everett} \& {Weisberg}}{{Everett} \&
  {Weisberg}}{2001}]{2001ApJ...553..341E}
{Everett}, J.~E.,  \& {Weisberg}, J.~M. 2001, \apj, 553, 341

\bibitem[\protect\citeauthoryear{{Ferri{\`e}re}}{{Ferri{\`e}re}}{2001}]{2001Rv%
MP...73.1031F}
{Ferri{\`e}re}, K.~M. 2001, Reviews of Modern Physics, 73, 1031

\bibitem[\protect\citeauthoryear{{Gaensler} et~al.}{{Gaensler}
  et~al.}{1998}]{1998ApJ...499L..69G}
{Gaensler}, B.~M., {Stappers}, B.~W., {Frail}, D.~A.,  \& {Johnston}, S. 1998,
  \apjl, 499, L69

\bibitem[\protect\citeauthoryear{{Gaensler} et~al.}{{Gaensler}
  et~al.}{2004}]{2004ApJ...616..383G}
{Gaensler}, B.~M., {van der Swaluw}, E., {Camilo}, F., {Kaspi}, V.~M.,
  {Baganoff}, F.~K., {Yusef-Zadeh}, F.,  \& {Manchester}, R.~N. 2004, \apj,
  616, 383

\bibitem[\protect\citeauthoryear{{Gotthelf}}{{Gotthelf}}{2003}]{2003ApJ...591.%
.361G}
{Gotthelf}, E.~V. 2003, \apj, 591, 361

\bibitem[\protect\citeauthoryear{{Harding} \& {Muslimov}}{{Harding} \&
  {Muslimov}}{2001}]{2001ApJ...556..987H}
{Harding}, A.~K.,  \& {Muslimov}, A.~G. 2001, \apj, 556, 987

\bibitem[\protect\citeauthoryear{{Harding} \& {Muslimov}}{{Harding} \&
  {Muslimov}}{2002}]{2002ApJ...568..862H}
{Harding}, A.~K.,  \& {Muslimov}, A.~G. 2002, \apj, 568, 862

\bibitem[\protect\citeauthoryear{{Heiles}}{{Heiles}}{2001}]{2001ApJ...551L.105%
H}
{Heiles}, C. 2001, \apjl, 551, L105

\bibitem[\protect\citeauthoryear{{Helfand}}{{Helfand}}{1983}]{1983IAUS..101..4%
71H}
{Helfand}, D.~J. 1983, in IAU Symp. 101: Supernova Remnants and their X-ray
  Emission, ed. J.~{Danziger} \& P.~{Gorenstein}, 471

\bibitem[\protect\citeauthoryear{{Hester} et~al.}{{Hester}
  et~al.}{2002}]{2002ApJ...577L..49H}
{Hester}, J.~J., et~al. 2002, \apjl, 577, L49

\bibitem[\protect\citeauthoryear{{Johnson} \& {Soderblom}}{{Johnson} \&
  {Soderblom}}{1987}]{1987AJ.....93..864J}
{Johnson}, D.~R.~H.,  \& {Soderblom}, D.~R. 1987, \aj, 93, 864

\bibitem[\protect\citeauthoryear{{Kalberla} et~al.}{{Kalberla}
  et~al.}{2005}]{2005A&A...440..775K}
{Kalberla}, P.~M.~W., {Burton}, W.~B., {Hartmann}, D., {Arnal}, E.~M.,
  {Bajaja}, E., {Morras}, R.,  \& {P{\"o}ppel}, W.~G.~L. 2005, \aap, 440, 775

\bibitem[\protect\citeauthoryear{{Kargaltsev} et~al.}{{Kargaltsev}
  et~al.}{2008}]{2008arXiv0802.2963K}
{Kargaltsev}, O., {Misanovic}, Z., {Pavlov}, G.~G., {Wong}, J.~A.,  \&
  {Garmire}, G.~P. 2008, ArXiv 0802.2963

\bibitem[\protect\citeauthoryear{{Kargaltsev} \& {Pavlov}}{{Kargaltsev} \&
  {Pavlov}}{2008}]{2008AIPC..983..171K}
{Kargaltsev}, O.,  \& {Pavlov}, G.~G. 2008, in American Institute of Physics
  Conference Series, Vol. 983, 40 Years of Pulsars: Millisecond Pulsars,
  Magnetars and More, 171

\bibitem[\protect\citeauthoryear{{Kargaltsev}, {Pavlov}, \&
  {Garmire}}{{Kargaltsev} et~al.}{2006}]{2006ApJ...636..406K}
{Kargaltsev}, O., {Pavlov}, G.~G.,  \& {Garmire}, G.~P. 2006, \apj, 636, 406

\bibitem[\protect\citeauthoryear{{Kargaltsev}, {Pavlov}, \&
  {Garmire}}{{Kargaltsev} et~al.}{2007}]{2007ApJ...660.1413K}
{Kargaltsev}, O., {Pavlov}, G.~G.,  \& {Garmire}, G.~P. 2007, \apj, 660, 1413

\bibitem[\protect\citeauthoryear{{Kargaltsev} et~al.}{{Kargaltsev}
  et~al.}{2006}]{2006HEAD....9.0757K}
{Kargaltsev}, O., {Pavlov}, G.~G., {Sanwal}, D., {Wong}, J.,  \& {Garmire},
  G.~P. 2006, in Bulletin of the American Astronomical Society, Vol.~38

\bibitem[\protect\citeauthoryear{{Kawai}, {Tamura}, \& {Saito}}{{Kawai}
  et~al.}{1998}]{1998AdSpR..21..213K}
{Kawai}, N., {Tamura}, K.,  \& {Saito}, Y. 1998, Advances in Space Research,
  21, 213

\bibitem[\protect\citeauthoryear{{Komissarov} \& {Lyubarsky}}{{Komissarov} \&
  {Lyubarsky}}{2003}]{2003MNRAS.344L..93K}
{Komissarov}, S.~S.,  \& {Lyubarsky}, Y.~E. 2003, \mnras, 344, L93

\bibitem[\protect\citeauthoryear{{McGowan} et~al.}{{McGowan}
  et~al.}{2007}]{2007Ap&SS.308..309M}
{McGowan}, K.~E., {Vestrand}, W.~T., {Kennea}, J.~A., {Zane}, S., {Cropper},
  M.,  \& {C{\'o}rdova}, F.~A. 2007, \apss, 308, 309

\bibitem[\protect\citeauthoryear{{Michel}}{{Michel}}{1991}]{1991tnsm.book.....%
M}
{Michel}, F.~C. 1991, {Theory of neutron star magnetospheres} (Univ. of Chicago
  Press, 1991, 533)

\bibitem[\protect\citeauthoryear{{Misanovic}, {Pavlov}, \&
  {Garmire}}{{Misanovic} et~al.}{2006}]{2006HEAD....9.0758M}
{Misanovic}, Z., {Pavlov}, G.~G.,  \& {Garmire}, G.~P. 2006, in Bulletin of the
  American Astronomical Society, Vol.~38

\bibitem[\protect\citeauthoryear{{Monet} et~al.}{{Monet}
  et~al.}{2003}]{2003AJ....125..984M}
{Monet}, D.~G., et~al. 2003, \aj, 125, 984

\bibitem[\protect\citeauthoryear{{Moon} et~al.}{{Moon}
  et~al.}{2004}]{2004ApJ...610L..33M}
{Moon}, D.-S., et~al. 2004, \apjl, 610, L33

\bibitem[\protect\citeauthoryear{{Mori} et~al.}{{Mori}
  et~al.}{2001}]{2001ASPC..251..576M}
{Mori}, K., {Tsunemi}, H., {Miyata}, E., {Baluta}, C.~J., {Burrows}, D.~N.,
  {Garmire}, G.~P.,  \& {Chartas}, G. 2001, in ASP Conf. Ser. 251: New Century
  of X-ray Astronomy, ed. H.~{Inoue} \& H.~{Kunieda}, 576

\bibitem[\protect\citeauthoryear{{Pavlov} et~al.}{{Pavlov}
  et~al.}{2007}]{2007ApJ...664.1072P}
{Pavlov}, G.~G., {Kargaltsev}, O., {Garmire}, G.~P.,  \& {Wolszczan}, A. 2007,
  \apj, 664, 1072

\bibitem[\protect\citeauthoryear{{Pavlov}, {Sanwal}, \& {Zavlin}}{{Pavlov}
  et~al.}{2006}]{2006ApJ...643.1146P}
{Pavlov}, G.~G., {Sanwal}, D.,  \& {Zavlin}, V.~E. 2006, \apj, 643, 1146

\bibitem[\protect\citeauthoryear{{Pavlov} et~al.}{{Pavlov}
  et~al.}{1995}]{1995lns..conf...71P}
{Pavlov}, G.~G., {Shibanov}, Y.~A., {Zavlin}, V.~E.,  \& {Meyer}, R.~D. 1995,
  in The Lives of the Neutron Stars. Proceedings of the NATO Advanced Study
  Institute on the Lives of the Neutron Stars, ed. M.~A. {Alpar},
  U.~{Kiziloglu}, \& J.~{van Paradijs}, 71

\bibitem[\protect\citeauthoryear{{Pavlov} et~al.}{{Pavlov}
  et~al.}{2003}]{2003ApJ...591.1157P}
{Pavlov}, G.~G., {Teter}, M.~A., {Kargaltsev}, O.,  \& {Sanwal}, D. 2003, \apj,
  591, 1157

\bibitem[\protect\citeauthoryear{{Pavlov} \& {Zavlin}}{{Pavlov} \&
  {Zavlin}}{1997}]{1997ApJ...490L..91P}
{Pavlov}, G.~G.,  \& {Zavlin}, V.~E. 1997, \apjl, 490, L91

\bibitem[\protect\citeauthoryear{{Romani} \& {Ng}}{{Romani} \&
  {Ng}}{2003}]{2003ApJ...585L..41R}
{Romani}, R.~W.,  \& {Ng}, C.-Y. 2003, \apjl, 585, L41

\bibitem[\protect\citeauthoryear{{Romanova}, {Chulsky}, \&
  {Lovelace}}{{Romanova} et~al.}{2005}]{2005ApJ...630.1020R}
{Romanova}, M.~M., {Chulsky}, G.~A.,  \& {Lovelace}, R.~V.~E. 2005, \apj, 630,
  1020

\bibitem[\protect\citeauthoryear{{S{\l}owikowska}, {Kuiper}, \&
  {Hermsen}}{{S{\l}owikowska} et~al.}{2005}]{2005A&A...434.1097S}
{S{\l}owikowska}, A., {Kuiper}, L.,  \& {Hermsen}, W. 2005, \aap, 434, 1097

\bibitem[\protect\citeauthoryear{{Stappers} et~al.}{{Stappers}
  et~al.}{2003}]{2003Sci...299.1372S}
{Stappers}, B.~W., {Gaensler}, B.~M., {Kaspi}, V.~M., {van der Klis}, M.,  \&
  {Lewin}, W.~H.~G. 2003, Science, 299, 1372

\bibitem[\protect\citeauthoryear{{Tepedelenl{\i}o{\v g}lu} \&
  {{\"O}gelman}}{{Tepedelenl{\i}o{\v g}lu} \&
  {{\"O}gelman}}{2005}]{2005ApJ...630L..57T}
{Tepedelenl{\i}o{\v g}lu}, E.,  \& {{\"O}gelman}, H. 2005, \apjl, 630, L57

\bibitem[\protect\citeauthoryear{{Tsunemi} et~al.}{{Tsunemi}
  et~al.}{2001}]{2001ApJ...554..496T}
{Tsunemi}, H., {Mori}, K., {Miyata}, E., {Baluta}, C., {Burrows}, D.~N.,
  {Garmire}, G.~P.,  \& {Chartas}, G. 2001, \apj, 554, 496

\bibitem[\protect\citeauthoryear{{van der Swaluw}}{{van der
  Swaluw}}{2003}]{2003A&A...404..939V}
{van der Swaluw}, E. 2003, \aap, 404, 939

\bibitem[\protect\citeauthoryear{{Vigelius} et~al.}{{Vigelius}
  et~al.}{2007}]{2007MNRAS.374..793V}
{Vigelius}, M., {Melatos}, A., {Chatterjee}, S., {Gaensler}, B.~M.,  \&
  {Ghavamian}, P. 2007, \mnras, 374, 793

\bibitem[\protect\citeauthoryear{{Wang} \& {Halpern}}{{Wang} \&
  {Halpern}}{1997}]{1997ApJ...482L.159W}
{Wang}, F.~Y.-H.,  \& {Halpern}, J.~P. 1997, \apjl, 482, L159

\bibitem[\protect\citeauthoryear{{Wang}, {Li}, \& {Begelman}}{{Wang}
  et~al.}{1993}]{1993Natur.364..127W}
{Wang}, Q.~D., {Li}, Z.-Y.,  \& {Begelman}, M.~C. 1993, \nat, 364, 127

\bibitem[\protect\citeauthoryear{{Weisskopf} et~al.}{{Weisskopf}
  et~al.}{2000}]{2000ApJ...536L..81W}
{Weisskopf}, M.~C., et~al. 2000, \apjl, 536, L81

\bibitem[\protect\citeauthoryear{{Yancopoulos}, {Hamilton}, \&
  {Helfand}}{{Yancopoulos} et~al.}{1994}]{1994ApJ...429..832Y}
{Yancopoulos}, S., {Hamilton}, T.~T.,  \& {Helfand}, D.~J. 1994, \apj, 429, 832

\bibitem[\protect\citeauthoryear{{Zavlin} \& {Pavlov}}{{Zavlin} \&
  {Pavlov}}{2004}]{2004ApJ...616..452Z}
{Zavlin}, V.~E.,  \& {Pavlov}, G.~G. 2004, \apj, 616, 452

\bibitem[\protect\citeauthoryear{{Zavlin}, {Shibanov}, \& {Pavlov}}{{Zavlin}
  et~al.}{1995}]{1995AstL...21..149Z}
{Zavlin}, V.~E., {Shibanov}, Y.~A.,  \& {Pavlov}, G.~G. 1995, Astronomy
  Letters, 21, 149

\bibitem[\protect\citeauthoryear{{Zhang}, {Sanwal}, \& {Pavlov}}{{Zhang}
  et~al.}{2005}]{2005ApJ...624L.109Z}
{Zhang}, B., {Sanwal}, D.,  \& {Pavlov}, G.~G. 2005, \apjl, 624, L109

\end{thebibliography}

\clearpage

\rotate{
\begin{table}
\scriptsize
\caption[]{{\sl Chandra} and {\sl XMM-Newton} observations of B1929 }
\begin{center}
\begin{tabular}{cccc}
\hline\hline\\
Instrument & Obs ID & Date & Exposure  \\
\hline\\
EPIC & 0113051301 & 10 Nov 2003 & 10.5 / 10.5 / 7.3 \\
EPIC & 0113051401 & 27 Apr 2004 & 15.2 / 16.4 / 11.0 \\
EPIC & 0113051501 & 29 Apr 2004 &  7.3 / 7.6 / 10.3 \\
ACIS & 6657 & 4 Dec 2005 & 20.9 \\
ACIS & 7230 & 28 May 2006 & 24.6 \\ 
\hline\\
\end{tabular}
\end{center}
\tablecomments{
The good-time exposures are given in ks. The first, second, and third
 EPIC exposures are for MOS1, MOS2,
and PN, respectively. The effective PN exposure in each observation is
 $\sim$70\% of the total PN 
good-time exposure because of the reduced efficiency of the small window mode.
}
\label{table-observations}
\end{table}
}

\clearpage

\rotate{
\begin{table}
\scriptsize
\caption[]{Background-subtracted counts and surface brightness of 
PWN regions in {\sl Chandra} observations 6657, 7230, and combined data}
\begin{center}
\begin{tabular}{ccccccc}
\hline\hline\\
Region &   &  1 (Southern Wing)  & 2 (Northern Wing) & 3 (Front) & 4 (Inner Blob) & 5 (Outer Blob)  \\
\hline \\
 Extraction area  &  & 31.5  & 46.7  &  77.9 & 157 & 1884 \\
arcsec$^2$ &  &  &   &   &  &  \\
\hline \\
Counts & 6657 & 18.9$\pm$4.6  & 6.9$\pm$3.2 & 12.7$\pm$4.2 & 23.5$\pm$5.8 & 
57$\pm$14  \\
(bkg-subtracted)    & 7230 & 15.3$\pm$4.1  & 17.3$\pm$4.5 & 3.7$\pm$2.8 & 14.4$\pm$4.8 & 
116$\pm$15  \\
 & comb. & 34.1$\pm$6.1   & 24.1$\pm$5.5  &  16.4$\pm$5.0 &
37.9$\pm$7.7 & 173$\pm$20  \\
&&&&&& \\
\hline \\
Bkg-subtracted & 6657 & 2.9$\pm$0.6  & 0.7$\pm$0.3 & 0.8$\pm$0.2 & 0.7$\pm$0.2 & 0.14$\pm$0.03   \\
 surface brightness  & 7230 & 2.0$\pm0.5$  & 1.5$\pm$0.4 & 0.2$\pm$0.1 & 0.4$\pm$0.1 & 0.25$\pm$0.03  \\
($10^{-5}$ cts\,s$^{-1}$\,arcsec$^{-2}$)  & comb. & 2.4$\pm$0.4  & 1.4$\pm$0.2  &  0.5$\pm$0.1 & 0.6$\pm$0.1 & 0.20$\pm$0.02  \\
 & & & & & & \\
\hline\\
\end{tabular}
\end{center}
\tablecomments{
The regions are shown in Fig.\ \ref{fig-combined-conv-smooth-image}.
 The errors represent statistical uncertainties at the 68\% confidence
level. 
 }
\label{table-regions}
\end{table}
}

\rotate{
\begin{table}
\scriptsize
\caption{
Fits to the spectrum of B1929
for the  {\sl Chandra} and {\sl XMM-Newton} observations.}
\begin{center}
\begin{tabular}{cccccccccc}
\hline\hline\\
Model & $N_{\rm H}$ & $\Gamma$ & PL Norm. & $kT$  & Radius\tablenotemark{a}  & $\chi_{\nu}^2$/d.o.f & Absorbed Flux  & Luminosity  \\
 &$10^{21}$ cm$^{-2}$ & & $10^{-5}$\,cm$^{-2}$\,s$^{-1}$\,keV$^{-1}$ & keV  & m &   & $10^{-13}$\,ergs\,cm$^{-2}$\,s$^{-1}$ & $10^{30}$\,ergs\,s$^{-1}$ \\
\hline \\
Chandra: &&&&&&&&\\
PL & $2.45^{+0.52}_{-0.48}$ & $2.94^{+0.25}_{-0.22}$ &  $7.99^{+1.76}_{-1.17}$ & $\cdots$ & $\cdots$ &
0.91/71 & $1.37^{+0.14}_{-0.27}$ & $6.26^{+2.66}_{-0.93}$ \\
PL+BB & $0.49^{+0.93}_{-0.49}$& $1.82^{+1.04}_{-0.56}$ & $1.91^{+3.86}_{-1.33}$ & $0.29^{+0.05}_{-0.06}$  & $33.8^{+46.3}_{-6.3}$ & 0.88/69 & $1.52^{+0.23}_{-0.37}$ & $4.66^{+0.86}_{-1.12}$ \\
XMM-Newton: & & & & & & & & \\
 PL & $2.22^{+0.37}_{-0.30}$ & $2.99^{+0.20}_{-0.15}$ & $8.96^{+1.32}_{-0.96}$ & $\cdots$ & $\cdots$ & 0.97/190 & $1.59^{+0.12}_{-0.25}$ & $7.08^{+2.18}_{-0.84}$ \\
 PL+BB & $0.05^{+0.11}_{-0.05}$& $1.63^{+0.86}_{-0.29}$ & $1.43^{+3.73}_{-0.42}$ & $0.30^{+0.02}_{-0.05}$ & $34.9^{+11.0}_{-4.9}$ & 0.94/188 & $1.82^{+0.13}_{-0.26}$ & $2.87^{+0.19}_{-0.39}$ \\
 combined: & & & & & & & & \\
 PL & $2.23^{+0.27}_{-0.27}$& $2.95^{+0.14}_{-0.13}$ &$8.43^{+0.91}_{-0.79}$ & $\cdots$ & $\cdots$ &  1.03/264 & $1.51^{+0.12}_{-0.24}$ & $6.62^{+1.98}_{-0.82}$ \\
 PL+BB & $0.17^{+0.23}_{-0.17}$& $1.73^{+0.46}_{-0.66}$ &$1.64^{+1.75}_{-0.28}$ &$0.30^{+0.02}_{-0.03}$ & $33.1^{+5.9}_{-4.6}$  &  0.98/262 & $1.75^{+0.11}_{-0.22}$ & $2.84^{+0.15}_{-0.22}$ \\
\hline \\
\end{tabular}
\end{center}
\tablecomments{The observed flux and unabsorbed luminosity, 
$L_X=4\pi d^2 F_X^{\rm unabs}$,
 are calculated for the 0.3--8\,keV energy band for {\sl Chandra}, and 0.3--10\,keV for {\sl XMM-Newton} and combined data. The listed uncertainties are at a 90\% confidence level determined for 2 interesting parameters.
}
\tablenotetext{a}{Radius of equivalent sphere for the BB component (see \$\,3.2).}
\label{table-spectrum}
\end{table}
}

\end{document}